\documentclass[11pt]{article}
\usepackage[margin=1in]{geometry}
\usepackage{amsmath,amsthm,amssymb,enumerate, graphicx,subcaption}
\usepackage{setspace}
\usepackage{thmtools}
\usepackage{thm-restate}
\usepackage[hidelinks]{hyperref}
\usepackage{cleveref,color}
\usepackage{natbib}
\usepackage[mathscr]{euscript}
\usepackage[shortlabels]{enumitem}
\usepackage{prodint}

\newif\ifproof
\prooffalse

\declaretheorem[name=Theorem]{theorem}

\declaretheorem[name=Corollary]{corollary}

\DeclareMathOperator*{\inprob}{\stackrel{P}{\longrightarrow}}

\DeclareMathOperator*{\indist}{\stackrel{d}{\longrightarrow}}

\newcommand{\fasterthan}{o_{P}}

\DeclareMathOperator*{\argmax}{argmax}

\renewcommand{\b}[1]{\mathbf{#1}}
\newcommand{\bs}[1]{\boldsymbol{#1}}
\newcommand{\s}[1]{\mathscr{#1}}
\renewcommand{\d}[1]{\mathbb{#1}}

\newcommand{\n}[1]{\mathrm{#1}}

\usepackage{natbib}

\allowdisplaybreaks

\title{Nonparametric maximum likelihood estimation\\under a likelihood ratio order}
\author{Ted Westling$^{1}$ \and Kevin J.\ Downes$^{2,3,4}$ \and Dylan S.\ Small$^{5}$}
\date{$^1$Department of Mathematics and Statistics, University of Massachusetts Amherst\\
$^2$Center for Pediatric Clinical Effectiveness, Children's Hospital of Philadelphia\\
$^3$Division of Infectious Diseases, Children's Hospital of Philadelphia\\
$^4$Department of Pediatrics, Perelman School of Medicine, University of Pennsylvania  \\
$^5$Department of Statistics, The Wharton School, University of Pennsylvania}
\begin{document}

\maketitle

\begin{abstract}
Comparison of two univariate distributions based on independent samples from them is a fundamental problem in statistics, with applications in a variety of scientific disciplines. In many situations, we might hypothesize that the two distributions are stochastically ordered, meaning that samples from one distribution tend to be larger than those from the other. One type of stochastic order is the likelihood ratio order, in which the ratio of the density functions of the two distributions is monotone non-decreasing. In this article, we derive and study the nonparametric maximum likelihood estimator of the individual distribution functions and the ratio of their densities under the likelihood ratio order. Our work applies to discrete distributions, continuous distributions, and mixed continuous-discrete distributions. We demonstrate convergence in distribution of the estimator in certain cases, and we illustrate our results using numerical experiments and an analysis of a  biomarker for predicting bacterial infection in children with systemic inflammatory response syndrome.
\end{abstract}

\doublespacing
\section{Introduction}

Comparing the distributions of two independent samples is a fundamental problem in statistics. Suppose that  $X_1, \dotsc, X_{n_1}$  and $Y_1, \dotsc, Y_{n_2}$ are independent real-valued samples with distribution functions $F_0$ and $G_0$, respectively. In many situations, we might hypothesize that $F_0$ and $G_0$ are \emph{stochastically ordered}, meaning intuitively that samples from $F_0$ tend to be larger than those from $G_0$. A particular type of stochastic order that arises in many applications is the \emph{likelihood ratio order}. Specifically, $G_0$ and $F_0$ satisfy a  likelihood ratio order if the density ratio $f_0/ g_0$ is monotone non-decreasing over the support $\s{G}_0$ of $G_0$, where $f_0 := dF_0 / d\eta$ and $g_0 := dG_0 /d\eta$ for some dominating measure $\eta$. For this reason, the likelihood ratio order is also called a \emph{density ratio order}. 

A likelihood ratio order can arise for a variety of scientific reasons \citep{beare2015tests, roosen2004testing, dykstra1995mle, yu2017ratio}. In the biomedical sciences and elsewhere, the ratio of two  density functions is an object of interest for describing the relative likelihood of a binary status indicator conditional on a covariate. If $D$ is a binary random variable, $Z$ is a scalar random variable, $F_0(z) = P(Z \leq z \mid D = 1)$, $G_0(z) = P(Z \leq z \mid D =0)$, and $H_0(z) := P(Z \leq z)$, then
\begin{equation} \frac{f_0(z)}{g_0(z)} =  \frac{[dF_0 / dH_0](z)}{[dG_0/ dH_0](z)} = \frac{P(D = 1 \mid Z =z)/ P(D=1)}{P(D = 0 \mid Z=z)/P(D=0)}\ .\label{eq:risk_ratio}\end{equation}
Therefore, the density ratio in this context may be interpreted as the relative odds of $D = 1$ given $Z = z$ to the overall odds of $D=1$. Since the transformation $x \mapsto x / (1-x)$ is strictly increasing, monotonicity of the density ratio is equivalent to monotonicity of $z \mapsto P(D = 1 \mid Z=z)$. One specific situation in which the representation given in~\eqref{eq:risk_ratio} is of scientific interest is biomarker evaluation, wherein $D$ represents infection status and $Z$ represents the value of a biomarker. Equation \eqref{eq:risk_ratio} implies that the ratio of the densities of biomarker values among infected patients to the same among uninfected patients can be interpreted as the odds ratio of infection given biomarker level relative to overall odds of infection. Monotonicity of the density ratio corresponds to the assumption that the conditional probability of infection given biomarker level increases with biomarker level, which is reasonable if the biomarker is actually predictive of disease.

In this article, we derive the nonparametric maximum likelihood estimators of $F_0$, $G_0$, and $\theta_0 = f_0 /g_0$ under the likelihood ratio order restriction and derive certain asymptotic properties of these estimators, including consistency and convergence in distribution. In particular, we use a connection between estimation of $\theta_0$ and the classical isotonic regression problem with a binary outcome, which both simplifies the derivation of large-sample results and suggests that existing inference methods for the isotonic regression problem can be used to perform inference for $\theta_0$ as well. Our results generalize those of \cite{dykstra1995mle}, who derived the maximum likelihood estimator of $F_0$ and $G_0$ under a likelihood ratio order in the special case where $F_0$ and $G_0$ are discrete distributions. We will illustrate our results using numerical experiments and an analysis of a  biomarker for predicting bacterial infection in children with systemic inflammatory response syndrome.

Recently, \cite{yu2017ratio} considered estimation of a monotone density ratio and the individual density functions by maximizing a smoothed likelihood function, and demonstrated certain asymptotic properties of their estimator.  \cite{yu2017ratio} considered maximizing a smoothed likelihood rather than maximizing the likelihood directly because the maximum likelihood estimator of the individual densities does not exist. In contrast, we will show that, using a definition of the likelihood ratio ordered model based on convexity of the ordinal dominance curve, a well-defined nonparametric maximum likelihood estimator of the monotone density ratio function and the individual distribution functions (rather than the density functions) does exist. Furthermore, unlike the smoothed estimator, the derivation of the maximum likelihood estimator does not require selection of a bandwidth or any other tuning parameter, and does not rely on the existence of Lebesgue density functions.

Additional relevant references include: \cite{lehmann1992orderings} and \cite{shaked2007stochastic}, which contain more examples and details regarding stochastic orders, \cite{carolan2005test} and \cite{beare2015tests}, which studied tests of the likelihood ratio order, and  \cite{rojo1991npmle}, \cite{rojo1993uniform}, \cite{mukerjee1996estimation}, \cite{arcones2000uniform}, \cite{davidov2012inference}, and \cite{tang2017ordering}, which considered testing and estimation under other stochastic orders.

\section{Likelihood ratio orders}

We observe two independent real-valued samples $X_1, \dotsc, X_{n_1}$ and $Y_1$, $\dotsc,$ $Y_{n_2}$ with distribution functions $F_0$ and $G_0$, respectively.  We define $\s{F}_0$ as the support of $F_0$ and $\s{G}_0$ as the support of $G_0$. We denote $n := n_1 + n_2$, and by $F_n$ and $G_n$ the empirical distribution functions of $X_1, \dotsc, X_{n_1}$ and $Y_1, \dotsc, Y_{n_2}$, respectively. We define $x_1 < \cdots < x_{m_1}$ as the unique values of $X_1, \dotsc, X_{n_1}$, $y_1 < \cdots < y_{m_2}$ as the unique values of $Y_1, \dotsc, Y_{n_2}$, and $z_1 < z_2 < \cdots < z_m$ as the unique values of $(X_1, \dotsc, X_{n_1}, Y_1, \dotsc, Y_{n_2})$. Throughout, we assume that $n$ is fixed, but that $n_1$ is drawn from a Binomial$(n,\pi_0)$ distribution for some $\pi_0 \in (0,1)$.

We let $\s{D}$ be the space of distribution functions on $\d{R}$; i.e.\ all non-decreasing, c\'{a}dl\`{a}g functions $H$ such that $\lim_{x \to -\infty} H(x) = 0$ and $\lim_{x \to \infty} H(x) = 1$. For any nondecreasing function $h : \d{R} \to \d{R}$, we define its \emph{generalized-inverse} $h^-$ pointwise as $h^-(u) := \inf\{x : h(x) \geq u\}$. When $h \in \s{D}$, $h^-$ is called the \emph{quantile function} of $h$. For any interval $I \subseteq \d{R}$ and any function $h : I \to \d{R}$, we define the \emph{greatest convex minorant} (GCM) of $h$ on $I$, denoted $\n{GCM}_I(h) : I \to\overline{\d{R}}$, for $\overline{\d{R}}$ the extended real line, as the pointwise supremum of all convex functions on $I$ bounded above by $h$. The least concave majorant operator is defined analogously. We say a function $H$ is \emph{convex over a set $\s{S} \subseteq \d{R}$} if for every $x, y \in \s{S}$ and $\lambda \in [0,1]$ such that $\lambda x + (1- \lambda) y \in \s{S}$, $H(\lambda x + (1-\lambda) y) \leq \lambda H(x) + (1-\lambda) H(y)$. We also define $\partial_-$ as the left derivative operator for a left differentiable function and $\n{Im}(h) := \{h(x) : x \in \s{S}\}$ as the image of a function $h$ defined on a domain $\s{S}$.

The unrestricted nonparametric model for the pair $(F, G)$ of distribution functions of the observed data is $\s{M}_{NP} := \s{D}^2$. As mentioned in the introduction, the likelihood ratio order can be defined as the ratio of the density functions $f_0$ and $g_0$ of $F_0$ and $G_0$ with respect to some dominating measure $\eta$ being non-decreasing. By varying the dominating measure, both discrete and continuous distributions can be handled this way. However, as noted by \cite{yu2017ratio}, this definition does not lend itself to the derivation of a maximum likelihood estimator, since the likelihood defined through the densities can be made arbitrarily large. Instead, other authors have defined the likelihood ratio order as convexity of the \emph{ordinal dominance curve}, defined as $t \mapsto R_{F,G}(t) := F \circ G^-(t)$ for $t \in [0,1]$  \citep{bamber1975odc, hsieh1996odc}.  \cite{lehmann1992orderings} demonstrated the equivalence of this definition to that using the density functions in the special case that $F$ and $G$ are strictly increasing and continuous on their supports, which were assumed to be intervals. Alternatively, \cite{shaked2007stochastic} defined the likelihood ratio order as $F(A)G(B) \leq F(B) G(A)$ for all measurable sets $A, B \subseteq \d{R}$ with $A \leq B$, where $F(A) := \int_A \, dF$ (with some abuse of notation) and $A \leq B$ means that $a \leq b$ for all $a \in A$ and $b \in B$.

In Theorem~\ref{thm:lik_ratio_order} below, we consolidate and generalize existing results connecting these different definitions of the likelihood ratio order.
\begin{theorem}\label{thm:lik_ratio_order}
If $F \ll G$ and $\nu := dF / dG$ is continuous on the support $\s{G}$ of $G$, then (1) the following are equivalent: $R_{F,G}$ is convex on $\n{Im}(G)$, $\nu$ is non-decreasing on $\s{G}$, and $F(A) G(B) \leq F(B) G(A)$ for all measurable sets $A \leq B$; and (2) if $\nu$ is non-decreasing on $\s{G}$ then $\nu(x) = \partial_-\n{GCM}_{[0,1]}(R_{F,G}) \circ G(x)$ for all  $x \in \s{G}$.
 \end{theorem}
To our knowledge, Theorem~\ref{thm:lik_ratio_order} is the most general result to-date connecting the three definitions of the likelihood ratio ordered model. We note that the three definitions may not be equivalent when $F$ is not dominated by $G$ or $\nu$ is not continuous. For instance, in the proof of Theorem~\ref{thm:lik_ratio_order} part (1), we only use the assumption that $\nu$ is continuous on $\s{G}$ to show that $R_{F,G}$ is convex on $\n{Im}(G)$ implies that $\nu$ is non-decreasing, but we show that if $\nu$ is non-decreasing, then $F(A) G(B) \leq F(B) G(A)$ for all $A \leq B$ (the definition used in \citealp{shaked2007stochastic}) regardless of whether $\nu$ is continuous. Additionally, we show that $F(A) G(B) \leq F(B) G(A)$ for all $A = (a_1, b_1] \leq B = (b_1, b_2]$ implies that $R_{F,G}$ is convex on $\n{Im}(G)$ regardless of whether $F \ll G$ or $\nu$ is continuous. However, to show the converse, we use $F \ll G$. For a simple counterexample when $F$ is not dominated by $G$, consider $F(\{a\}) = 1$ and $G(\{b\}) = 1$, where $a < b$. Then $R_{F,G}(u) = I(u > 0)$ for $u \in [0,1]$, which is convex on $\n{Im}(G) = \{0,1\}$, but $1 = F(A)G(B) > F(B) G(A) = 0$ for $A = \{a\} \leq \{b\} = B$. Finally, when $F \ll G$ but $dF / dG$ is not continuous on $\s{G}$, whether $R_{F,G}$ being convex on $\n{Im}(G)$ implies that $F(A) G(B) \leq F(B) G(A)$ for \emph{all} measurable sets $A \leq B$, or even all such Borel sets, is unclear to us.

Throughout the remainder of the article, we say $(F, G) \in \s{M}_{NP}$ satisfy a likelihood ratio order, and write $G \leq_{LR} F$ if $R_{F,G}$ is convex on $\n{Im}(G)$. We then define the likelihood ratio ordered model $\s{M}_{LR}$ as all $(F, G) \in \s{M}_{NP}$ such that $G \leq_{LR} F$.  For any $(F, G) \in \s{M}_{NP}$, we further define $\theta : \s{M}_{NP} \to \bs\Theta$ as $\theta_{F, G} := \partial_-\n{GCM}_{[0,1]}(R_{F,G}) \circ G$, where $\bs\Theta$ is defined as the set of non-negative, non-decreasing functions on $\d{R}$. We note that this definition allows for the possibility that $F$ is not dominated by $G$, but by Theorem~\ref{thm:lik_ratio_order}, for all $(F,G) \in \s{M}_{LR}$ such that $F \ll G$ and $dF / dG$ is continuous on $\s{G}$, $\theta_{F, G} =  dF / dG$ on $\s{G}$. We define $\theta_0 := \theta_{F_0, G_0}$. 

In the context of the likelihood ratio order, many existing works either assume that $F_0$ and $G_0$ are discrete (e.g.\ \citealp{dykstra1995mle}) or that $F_0$ and $G_0$ are continuous (e.g.\ \citealp{lehmann1992orderings, yu2017ratio}). In the discrete setting, if $F_0$ and $G_0$ are discrete distributions with common support and mass functions $\Delta F_0$ and $\Delta G_0$ such that $(F_0, G_0) \in \s{M}_{LR}$, then $\theta_0 = \Delta F_0 / \Delta G_0$ on $\s{G}_0$. Alternatively, if $F_0$ and $G_0$ both possess Lebesgue density functions $f_0$ and $g_0$ and $(F_0, G_0) \in \s{M}_{LR}$, then $\theta_0 = f_0 / g_0$ on $\s{G}_0$. However, for the purpose of deriving a maximum likelihood estimator, we will demonstrate that these two cases do not need to be treated separately. Furthermore, in some applied settings, $F_0$ and $G_0$ are neither discrete nor continuous, but rather a mixture of discrete and continuous components, and we will derive results that apply in these situations as well. For instance, exposures that are bounded below may have positive mass at their lower boundary, and be continuous thereafter. Many biomarkers exhibit this property. Similarly, some measurements are ``clumpy", exhibiting positive mass at integers or other ``round" numbers due to the measurement process, but also possessing positive Lebesgue density between such points. In all cases, $\theta_0$ has a meaningful interpretation as the ratio of the conditional odds of a sample being from the distribution $F_0$ to the unconditional odds of a sample being from $F_0$.

\section{Estimation under a likelihood ratio order}\label{sec:estimation}

\subsection{Maximum likelihood estimator}

The pair $(F_0, G_0)$ determines the joint distribution of the observed data. Defining the nonparametric likelihood of the observed data as
$L_n(F, G) :=  \prod_{i=1}^{n_1} \left[ F(X_i) - F(X_i-) \right]\prod_{j=1}^{n_2} \left[ G(Y_j) - G(Y_j-)  \right]$, the nonparametric maximum likelihood estimator of $(F_0, G_0)$, i.e.\ in the model $\s{M}_{NP}$, is $(F_n, G_n)$ for $F_n$ the empirical distribution function of $X_1, \dotsc, X_{n_1}$, and $G_n$ the same of $Y_1, \dotsc, Y_{n_2}$. This suggests taking as an estimator of $\theta_0$ the plug-in estimator $\theta_n := \theta_{F_n, G_n} = \partial_-\n{GCM}_{[0,1]}(F_n \circ G_n^-) \circ G_n$. The function $F_n \circ G_n^-$ is known as the \emph{empirical ordinal dominance curve}, and is properties  were studied by \cite{hsieh1996odc}.

In this section, we demonstrate, amongst other results, that $\theta_n$ is the maximum likelihood estimator of $\theta_0$ in the likelihood ratio ordered model $\s{M}_{LR}$. 
A maximum likelihood estimator of $(F_0, G_0)$ in $\s{M}_{LR}$ is defined as $(F_n^*, G_n^*) \in \argmax_{(F, G) \in \s{M}_{LR}} L_n(F,G)$, and a maximum likelihood estimator of $\theta_0$ is defined as $\theta_n^* := \theta_{F_n^*, G_n^*}$. 

We define $H_n(z) :=\pi_n F_n(z) + (1-\pi_n) G_n(z)$ as the empirical distribution of the combined sample $X_1, \dotsc, X_{n_1}, Y_1, \dotsc, Y_{n_2}$, and $h_k := H_n(y_k)$ for $k = 1, \dotsc, m_2$. Our first result characterizes $(F_n^*, G_n^*)$.
\begin{theorem}\label{thm:mle}
Let  $A_k^*$ be the value at $h_k$ of the GCM over $[0,h_{m_2}]$ of $\{ (h_k, F_n(y_k)) : k=0, \dotsc, m_2\}$ and $B_k^*$ be the value at $h_k$ of the LCM over $[0, h_{m_2}]$ of $\left\{ \left(h_k, G_n(y_k)\right) :  k=0, \dotsc, m_2\right\}$. Then $G_n^*$ is a right-continuous step function with jumps at $y_1, \dotsc, y_{m_2}$ with $G_n^*(y_k) = B_k^*$ and $F_n^*$ is given by a right-continuous step function with jumps at $z_1, \dotsc, z_m$, where $F_n^*(y_k) = A_k^*$, and for any $x_i$ such that $y_{j-1} < x_i \leq y_j$, where $y_0 := -\infty$, the mass of $F_n^*$ at $x_i$ is given by 
\[ F_n^*(x_i) - F_n^*(x_i-) = \left[F_n^*(y_j) - F_n^*(y_{j-1})\right] \frac{F_n(x_i) - F_n(x_i-)}{F_n(y_j) - F_n(y_{j-1})} \ . \]
For any $x_i$ such that $y_{m_2} < x_i$, the mass of $F_n^*$ at $x_i$ is given by 
\[ F_n^*(x_i) - F_n^*(x_i-) = \left[1- F_n^*(y_{m_2})\right] \frac{F_n(x_i) - F_n(x_i-)}{1 - F_n(y_{m_2})} \ . \]
We also note that $F_n^*(y_k) = \n{GCM}_{[0,h_{m_2}]} (F_n \circ H_n^{-})(H_n(y_k))$ and $G_n^*(y_k) =  \n{LCM}_{[0,h_{m_2}]} (G_n \circ H_n^{-})(H_n(y_k))$. 
\end{theorem}

A proof of Theorem~\ref{thm:mle}, and proofs of all other theorems, are provided in Supplementary Material. We note that $F_n^*$ necessarily has jumps at all $x_i$ and at all $y_j$ such that $y_j \geq x_1$, and $G_n^*$ has jumps at all $y_j$. We also note that if there are $j$ such that no $x_i \in (y_j, y_{j+1}]$ but $F_n^*(y_{j}) > F_n^*(y_{j-1})$, then there are infinitely many maximizers $F_n^*$ because any $F_n^*$ that assigns mass $F_n^*(y_{j}) - F_n^*(y_{j-1})$ to the interval $(y_j, y_{j+1}]$ yields the same likelihood and satisfies the constraints. In these cases, for the sake of uniqueness, we will put mass $F_n^*(y_{j}) - F_n^*(y_{j-1})$ at the point $y_{j+1}$. 

Theorem~\ref{thm:mle}  implies the following result characterizing $\theta_n^*$.
\begin{corollary}\label{cor:theta_mle}
The points $\{(G_n^*(y_k), F_n^*(y_k)): k=1, \dotsc, m_2\}$ lie on the GCM over $[0,1]$ of the empirical ordinal dominance curve 
\[\left\{ (G_n(y_j), F_n(y_j)) : k=0,\dotsc, m_2\right\},\] where $y_0:= -\infty$. Specifically, if $\left\{ \left(h_{j_k}, F_n(y_{j_k})\right) : k=0, \dotsc, K\right\}$ are the vertices of the GCM of $\left\{ \left(h_k, F_n(y_k)\right) : k=0, \dotsc, m_2\right\}$, then $(G_n(y_{j_k}), F_n(y_{j_k})) : k = 0, \dotsc, K\}$ are the vertices of the GCM of the empirical ordinal dominance curve. Therefore, $\theta_n^* := \theta_{F_n^*, G_n^*}$ is equal to $\theta_n := \theta_{F_n, G_n}$.
\end{corollary}

Theorem~\ref{thm:mle} bears resemblance to, but is different than, Theorem~2.1 of \cite{dykstra1995mle}, which characterized the maximum likelihood estimator under a likelihood ratio order in the discrete case. Here, we perform the maximization over all pairs of univariate distribution functions $(F, G)$ such that $R_{F,G} = F \circ G^-$ is convex on the support of $G$, whereas Theorem~2.1 of \cite{dykstra1995mle} performed the maximization over $(F,G)$ with support contained in $\{z_1, \dotsc, z_{m}\}$ and such that $[\Delta F(z_j)] / [\Delta G(z_j)]$ is nondecreasing. The first set is  strictly larger than the second, which results in possibly different maximum likelihood estimators. In particular, our maximum likelihood estimator $G_n^*$ is only supported on $y_1, \dotsc, y_{m_2}$, whereas the maximum likelihood estimator of $G_0$ derived by \cite{dykstra1995mle} may have support on $x_j$ that are not equal to any $y_1, \dotsc, y_{m_2}$. This difference makes sense in the context of our respective problem formulations:  \cite{dykstra1995mle} assumed that the supports of $F_0$ and $G_0$ are subsets of $\{z_1, \dotsc, z_{m}\}$, while we do not assume the supports are known \emph{a priori}. In Supplementary Material, we illustrate the use of Theorem~\ref{thm:mle} using hypothetical data in which our maximum likelihood estimators $F_n^*$ and $G_n^*$ are different from those of \cite{dykstra1995mle}.

\subsection{Representation as a transformation of isotonic regression}\label{sec:iso}

\cite{dykstra1995mle} and \cite{carolan2005test} provided representations of the maximum likelihood estimators of $F_0$ and $G_0$ in terms of isotonic regression in the discrete and continuous cases, respectively. Here, we show that $\theta_n^*$ can also be represented as a transformation of an isotonic regression, which aids in deriving its asymptotic properties. We let $D_1, \dotsc, D_n$ be independent Bernoulli random variables with common probability $\pi_0$ and such that $n_1 = \sum_{i=1}^n D_i$. Letting $j_1, \dotsc, j_{n_1}$ be the indices such that $D_{j_i} = 1$ for each $i$, we then define $Z_{j_i} := X_i$ for each $i = 1, \dotsc, n_1$. Similarly, letting $k_1, \dotsc, k_{n_2}$ be the indices such that $D_{k_i} = 0$ for each $i$, we define $Z_{k_i}:= Y_i$ for each $i = 1, \dotsc, n_2$.  Defining the data unit $\b{O}_i := (Z_i, D_i)$, observing the independent samples $X_1, \dotsc, X_{n_1}$ from $F_0$ and $Y_1, \dotsc, Y_{n_2}$ from $G_0$ is then equivalent to observing independent observations $\b{O}_1, \dotsc, \b{O}_n$ from $P_0$, where $P_0$ satisfies
\[ P_0(Z \leq z, D = d)  = d\pi_0 F_0(z) + (1-d)(1-\pi_0)G_0(z)\ .\]
Thus, $Z_1, \dotsc, Z_n$ represent the pooled values of $X_1, \dotsc, X_{n_1}, Y_1, \dotsc, Y_{n_2}$, and each $D_i$ represents an indicator that $Z_i$ corresponds to a sample from $F_0$. Furthermore, $F_0(z) = P_0(Z \leq z \mid D = 1)$, $G_0(z) = P_0(Z \leq z \mid D = 0)$, and $\pi_0 :=  P_0(D = 1)$. Estimating $\theta_0$ given the independent samples $X_1, \dotsc, X_{n_1}$ and $Y_1, \dotsc, Y_{n_2}$ is therefore equivalent to estimating $\theta_0$ given independent observations $\b{O}_1, \dotsc, \b{O}_n$ from $P_0$, where $n_1 := \sum_{i=1}^n D_i$.

The benefit to the above reframing of the problem is that $\theta_0$, $F_0$, and $G_0$ can then be written as transformations of $P_0$. First, we have that $\theta_0(z) = T(\mu_0(z)) / T(\pi_0)$, where $\mu_0(z) := P_0(D = 1 \mid Z = z)$ and $T: [0,1) \to \d{R}^+$ is the odds transformation,  defined as $T(\mu) := \mu / (1-\mu)$.  Since $T$ is strictly increasing, $\theta_0$ is monotone if and only if $\mu_0$ is. Since the maximum likelihood estimator of $\mu_0$ under the assumption that $\mu_0$ is non-decreasing is given by the isotonic regression $\mu_n^*$ of $D_1, \dotsc, D_n$ on $Z_1, \dotsc, Z_n$, and the maximum likelihood estimator of $\pi_0$ is given by $\pi_n$, the maximum likelihood estimator of $\theta_0(z)$ is then given by $T(\mu_n^*(z)) / T(\pi_n)$. It is straightforward to see that this form of the maximum likelihood estimator is equivalent to the forms given above. In the next section, we will utilize this form of $\theta_n^*$ to derive its asymptotic properties  and to construct asymptotic confidence intervals.

\section{Asymptotic results}\label{sec:asymptotic}

\subsection{Discrete distributions}\label{sec:asy_discrete}

We first consider the situation where $G_0$ has finite support $\s{G}_0$ and $\theta_0$ is strictly increasing on $\s{G}_0$. The next result demonstrates that in this case, $F_n^*$ and $G_n^*$ are asymptotically equivalent to $F_n$ and $G_n$, respectively, and $\theta_n^*$ is asymptotically equivalent to the ratio of the empirical masses on the support of $G_0$.
\begin{theorem}[Discrete distributions]
\label{thm:discrete}
Suppose that the support $\s{G}$ of $G_0$ is a finite set $\{y_1 < y_2 < \cdots < y_{m_2}\}$ and that $[F_0(y_j) - F_0(y_{j-1})]/ \Delta G_0(y_j) < [F_0(y_{j+1}) - F_0(y_{j})]/ \Delta G_0(y_{j+1}) $ for $j = 1, \dotsc, m_2-1$, where $y_0 := -\infty$. Then $F_n^* = F_n$ and $G_n^* = G_n$ with probability tending to one, so that with probability tending to one $\theta_n^*$ is a left-continuous step function with jumps at $y_1, \dotsc, y_{m_2-1}$ and $\theta_n^*(y_j) = [F_n(y_j) - F_n(y_{j-1})] / \Delta G_n(y_j)$ and $\theta_n^*(z) = 0$ for $z < y_1$. As a result, $n^{1/2} [\theta_n^*(y_j) - \theta_0(y_j)] \indist N(0, \sigma_0^2(y_j))$ for
\[\sigma_0^2(y_j) := \theta_0(y_j) \frac{\pi_0 F_{0,j} + (1-\pi_0) \Delta G_0(y_j) -F_{0,j} \Delta G_0(y_j)}{\pi_0(1-\pi_0) [\Delta G_0(y_j)]^2},\]
where $F_{0,j} := F_0(y_j) - F_0(y_{j-1})$. 
\end{theorem}
We note that the above result does not require that $F_0$ be discrete as well, or be dominated by $G_0$. If $F_0$ is dominated by $G_0$, then $\theta_0 = \Delta F_0 / \Delta G_0$ corresponds to the ratio of the mass functions.

\subsection{Continuous distributions}\label{sec:asy_cont}

Now we address the situation where $F_0$ and $G_0$ are both absolutely continuous on $\s{G}_0$ and $\theta_0$, which now corresponds to the ratio $f_0/g_0$ of the density functions, is strictly increasing. We first consider the large-sample behavior of $F_n^*$ and $G_n^*$. 
\begin{theorem}
\label{thm:cont_dist_fns}
Suppose that $G_0$ is supported on a bounded interval $[a, b] \subset \d{R}$, that $F_0$ and $G_0$ possess continuous density functions $f_0$ and $g_0$ on $[a,b]$ such that $f_0 / g_0$ is strictly increasing on $[a,b]$, and $g_0(z) \geq \kappa > 0$ on $[a,b]$. Then  $\| G_n^* - G_n\|_{\infty} = \fasterthan(n^{-1/2})$ and $\| F_n^* - F_n\|_{\infty} = \fasterthan(n^{-1/2})$.
\end{theorem}
Theorem~\ref{thm:cont_dist_fns} demonstrates that when $\theta_0$ is strictly increasing, the maximum likelihood estimators of the individual distribution functions are asymptotically equivalent to the empirical distribution functions at the rate $n^{-1/2}$, and hence possess the same limit distributions as the empirical distribution functions. This result is proved using the functional delta method and the results of \cite{beare2017weak}, who demonstrated that the LCM operation is a directionally Hadamard differentiable mapping at any concave function.

We now turn to large-sample results for $\theta_n^*$ at points $z$ where both $F_0$ and $G_0$ possess Lebesgue density functions $f_0$ and $g_0$, respectively. First, consistency of $\mu_n^*$ implies consistency of $\theta_n^*$.
\begin{theorem}[Consistency]
If $f_0$ is continuous at $x$, $g_0$ is continuous at $x$, and $g_0(x) > 0$, then $\theta_n^*(x) \inprob \theta_0(x)$. If $f_0$ and $g_0$ are uniformly continuous on $\s{G}_0$, then $\sup_{x \in I} | \theta_n^*(x) - \theta_0(x) | \inprob 0$ for any strict sub-interval $I \subsetneq \s{G}_0$. 
\label{thm:cons}
\end{theorem}

We recall that, at any $z$ such that $h_0 = \pi_0 f_0 + (1-\pi_0) g_0$ is positive and continuous in a neighborhood of $z$, $\mu_0(z) \in (0,1)$, and $\mu_0$ is continuously differentiable in a neighborhood of $z$, it holds that
\begin{equation} n^{1/3}\left[ \mu_n^*(z) - \mu_0(z)\right] \indist \left\{ 4 \mu_0'(z) \mu_0(z) [1-\mu_0(z)] h_0(z)^{-1} \right\}^{1/3} W \ ,\label{eq:mu_limit}\end{equation}
where $W$ follows \emph{Chernoff's distribution}, defined as the point of maximum of $Z(u) - u^2$ for $Z$ a two-sided standard Brownian motion originating from zero \citep{brunk1970regression, groene2014shape}. We can then use the delta-method to see that
\begin{align*}
n^{1/3}\left[\theta_n^*(z) - \theta_0(z)\right]
&\indist T(\pi_0) T'(\mu_0) \left\{ 4 \mu_0'(z) \mu_0(z) [1-\mu_0(z)] h_0(z)^{-1} \right\}^{1/3} W\ .
\end{align*}
The scale parameter in the above limit distribution is equal to $\left[4\kappa_0(z) \theta_0'(z) \right]^{1/3}$  for
\[ \kappa_0(z) :=\theta_0(z) \frac{\pi_0f_0(z)  + (1-\pi_0)g_0(z)}{\pi_0(1-\pi_0)g_0(z)^2} \ .\]
This yields the following result.
\begin{theorem}[Pointwise convergence in distribution]
\label{thm:dist}
Suppose that, in a neighborhood of $z$, $\theta_0$ is continuously differentiable with $\theta_0'(z) > 0$, and $f_0$ and $g_0$ are positive and continuous. Then
$n^{1/3} [ \theta_n^*(z) - \theta_0(z)] \indist \left[4\kappa_0(z) \theta_0'(z) \right]^{1/3} W.$
\end{theorem}

Theorem~\ref{thm:dist} reflects certain common tradeoffs in the monotonicity-constrained literature. Theorem~\ref{thm:dist} indicates that the non-smoothed estimator converges pointwise at the $n^{-1/3}$ rate.  In contrast, the smoothed estimator proposed by \cite{yu2017ratio} converges at the faster $n^{-2/5}$ rate, albeit under stronger smoothness assumptions. While \cite{yu2017ratio} did not propose a method for conducting inference, smoothed estimators typically possess an asymptotic bias that complicates the task of performing valid inference. In contrast, the limit distribution in Theorem~\ref{thm:dist} has mean zero, which we can use to construct asymptotically valid confidence intervals. Defining $\tau_n(z)$ as an estimator of $\tau_0(z) := \kappa_0(z) \theta_0'(z)$ and $q_\alpha$ the $1-\alpha/2$ quantile of $W$, a $100(1-\alpha)\%$ Wald-type confidence interval for $\theta_0(z)$ is given by 
$ \theta_n^*(z) \pm  [4\tau_n(z)/n]^{1/3} q_{1-\alpha/2}$.
If $\tau_n(z) \inprob \tau_0(z)$, then this interval has asymptotic coverage of $100(1-\alpha)\%$. The quantiles of $W$ were computed by \cite{groeneboom2001computing}, and in particular $q_{0.975} \approx 0.9982$.

In practice, we recommend an alternative method to constructing confidence intervals for $\theta_0(z)$. We recommend first constructing confidence intervals for $\mu_0(z)$ using either of two existing methods, then transforming these intervals into intervals for $\theta_0(z)$. Specifically, if $[\ell_n(z), u_n(z)]$ represents a $100(1-\alpha)\%$ confidence interval for $\mu_0(z)$, then we take $[ T(\ell_n(z)) / T(\pi_n),$ $T(u_n(z)) / T(\pi_n) ]$ as a $100(1-\alpha)\%$ confidence interval for $\theta_0(z)$. Two existing ways to construct $[\ell_n(z), u_n(z)]$ are Wald-type intervals with plug-in estimation of nuisance parameters and intervals based on likelihood ratio tests. The former intervals are analogous to the Wald-type interval, but based on the limit distribution for $n^{1/3}[ \mu_n^*(z) - \mu_0(z)]$ given in \eqref{eq:mu_limit}. Alternatively, confidence intervals obtained by inverting likelihood ratio tests, proposed first by \cite{banerjee2001ratio} and studied further by, e.g.\ \cite{banerjee2007response} and \cite{groeneboom2015nonparametric}, can be formed based on the limiting distribution of twice the log of the ratio of the likelihoods of the maximum likelihood estimator and a suitably constrained maximum likelihood estimator. Since this limiting distribution is pivotal, meaning it does not depend on any unknown features of the true distribution, this approach does not require estimating any unknown nuisance parameters. We therefore expect this method to have better finite-sample properties than intervals based on plug-in estimation of nuisance parameters.

\section{Numerical studies}\label{sec:numerical}

In Supplementary Material, we present results of two simulation studies in the cases where $F_0$ and $G_0$ are fully discrete and fully continuous. In short, these studies confirm the validity of our large-sample theory and demonstrate that the maximum likelihood estimator and various proposed methods of conducting inference perform well in both cases. Here, we present the results of a numerical study illustrating the behavior of $\theta_n^*$ when $F_0$ and $G_0$ are mixed discrete-continuous distributions. We note that our asymptotic results did not address the behavior of $\theta_n^*$ at mass points in mixed discrete-continuous distributions; to the best of our knowledge, no such results yet exist for monotone estimators. We use this numerical study to explore this important case.

We simulated $Y$ as a mixed discrete-continuous random variable with probability 1/9 each of being 0, 0.5 and 1, and probability 2/3 of being from the uniform distribution on $[0,1]$, and simulated $X$ as a mixed discrete-continuous random variable  with probabilities 1/18, 1/9, and 3/18 of being 0, 0.5, and 1, respectively and probability 2/3 of being from the density function $x \mapsto I_{[0,1]}(x) (0.5 + x)$. We then have that $\theta_0(x) = 0.5 + x$ for $x \in [0,1]$. We set $\pi_0 := 0.4$. For each combined sample size $n \in \{500, 1K, 5K, 10K\}$, we simulated $1000$ datasets, and in each dataset we computed the maximum likelihood estimator, the maximum smoothed likelihood estimator of \cite{yu2017ratio}, and the non-monotone estimator based on kernel density estimates for each $z \in \{0, 0.05, \dotsc, 0.95, 1\}$. We constructed confidence intervals at each $z$ using the transformed plug-in and likelihood ratio-based methods described in Section~\ref{sec:asy_cont}. To estimate the scale parameter in the limit distribution of $\mu_n^*(z)$ as defined in equation~\ref{eq:mu_limit}, we used the plug-in estimator $\mu_n^*(z)$ for $\mu_0(z)$ and estimated $\mu_0'(z) / h_0(z) = (\mu_0 \circ H_0^{-1})' \circ H_0(z)$ using the derivative of a local linear smoother of $\mu_n^* \circ H_n^{-}$ evaluated at $H_n(z)$.

In addition to the properties of the estimators listed above, we also investigated the properties of the general sample-splitting procedure proposed by \cite{banerjee2019divide}. Given a generic monotone estimator $\gamma_n$ of a monotone function $\gamma_0$ such that $n^{1/3}[ \gamma_n(z) - \gamma_0(z)] \indist G$ for $G$ a mean-zero distribution with finite variance, \cite{banerjee2019divide} proposed randomly splitting the sample into $m$ subsets of roughly equal size, computing monotone estimates $\gamma_{n,1}, \dotsc, \gamma_{n,m}$ in each subset, then defining $\bar\gamma_{n,m}(z) := \frac{1}{m} \sum_{j=1}^m \gamma_{n,j}(z)$. They demonstrated that if $m > 1$ is fixed, then under mild conditions $\bar\gamma_{n,m}(z)$ has strictly better asymptotic mean squared error than $\gamma_n(z)$, and that for moderate $m$,
$\bar\gamma_{n,m}(z) \pm \sigma_{n,m}(z)t_{1-\alpha/2, m-1} /\sqrt{m}$
forms an asymptotic $100(1-\alpha)\%$ confidence interval for $\gamma_0(z)$, where $\sigma_{n,m}^2(z) := \frac{1}{m-1} \sum_{j=1}^m [ \gamma_{n,j}(z) - \bar\gamma_{n,m}(z)]^2$ and $t_{1-\alpha/2, m-1}$ is the $100(1-\alpha/2)$ quantile of the $t$-distribution with $m-1$ degrees of freedom. Therefore, $\bar\gamma_{n,m}(z)$ is preferable to $\gamma_n(z)$ for two reasons: it has better asymptotic mean squared error, and asymptotically valid pointwise confidence intervals for $\gamma_0$ based on $\bar\gamma_{n,m}$ can be formed without estimating any nuisance parameters. They also studied the asymptotic properties of $\bar\gamma_{n,m_n}(z)$ when $m_n$ grows with $n$. In our simulation study, we considered the estimator $\bar\theta_{n,m}$ defined as $\bar\theta_{n,m}(z) := \frac{1}{m} \sum_{j=1}^m \theta_{n,j}^*(z)$, where $\theta_{n,j}^*$ is the maximum likelihood estimator in the $j$th subset, and the corresponding confidence intervals defined above. We only considered the situation where $m \in \{5,10\}$ is fixed with the sample size. 

\begin{figure}[h!]
\centering
\includegraphics[width=.49\linewidth]{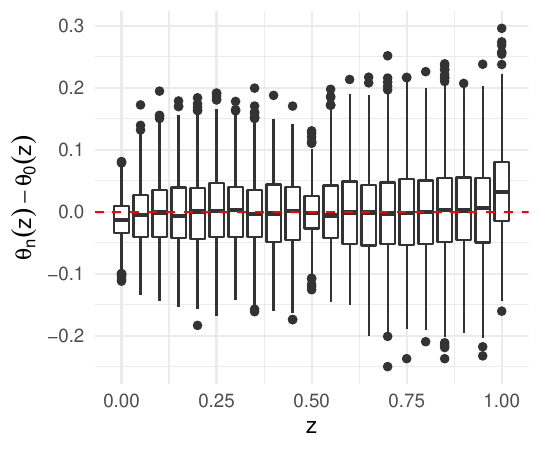}
\includegraphics[width=.49\linewidth]{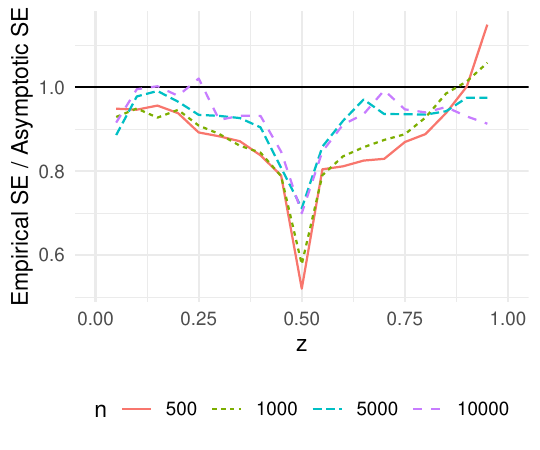}\par
\caption{Left: boxplots of $\theta_n^*(z) - \theta_0(z)$ with $n=10K$. Right: empirical standard errors of $r_n[\theta_n^*(z) - \theta_0(z)]$ divided by the limit theory-based counterparts for $z \in (0,1)$, where $r_n = n^{1/2}$ for $z =0.5$ and $r_n = n^{1/3}$ otherwise.}
\label{fig:sd_empirical_vs_true}
\end{figure}

We now turn to the results of the simulation study. The left panel of Figure~\ref{fig:sd_empirical_vs_true} displays the distribution of $\theta_n^*(z) - \theta_0(z)$ for $z \in [0,1]$ and $n=10K$. These distributions are approximately centered around 0 for $z \in (0,1)$, but not for $z \in \{0,1\}$. Hence, despite the positive mass at the boundaries, the maximum likelihood estimator does not appear to be consistent at the boundaries. This is a common problem among monotonicity-constrained estimators, and various correction procedures have been proposed and could be considered in this context (see, e.g.\ \citealp{woodroofe1993penalized, kulikov2006}).

The right panel of Figure~\ref{fig:sd_empirical_vs_true} displays the ratio of the standard deviation of $r_n[\theta_n^*(z) - \theta_0(z)]$ to the standard deviation of the asymptotic distributions derived in Section~\ref{sec:asymptotic} for $z \neq 0,1$. For $z = 0.5$, $r_n = n^{1/2}$ and the asymptotic distribution is that of the fully discrete case presented in Section~\ref{sec:asy_discrete}, though we note that the results presented in that section do not apply here due to the mixed discrete-continuous nature of $F_0$ and $G_0$ here. Otherwise, $r_n = n^{1/3}$ and the asymptotic distribution is that of the continuous case presented in Section~\ref{sec:asy_cont}. We see that, for $z \neq 0.5$, the empirical standard error approaches the asymptotic standard deviation as $n$ grows. However, for $z = 0.5$, the empirical standard error is converging to a limit that is strictly smaller than the asymptotic standard deviation. This suggests that, at points that have both positive mass and positive density in a neighborhood of the point, the maximum likelihood estimator gains efficiency from the positive density. In addition, points of continuity near the mass point also experience finite-sample efficiency gains.

\begin{figure}[h!]
\centering
\includegraphics[width=\linewidth]{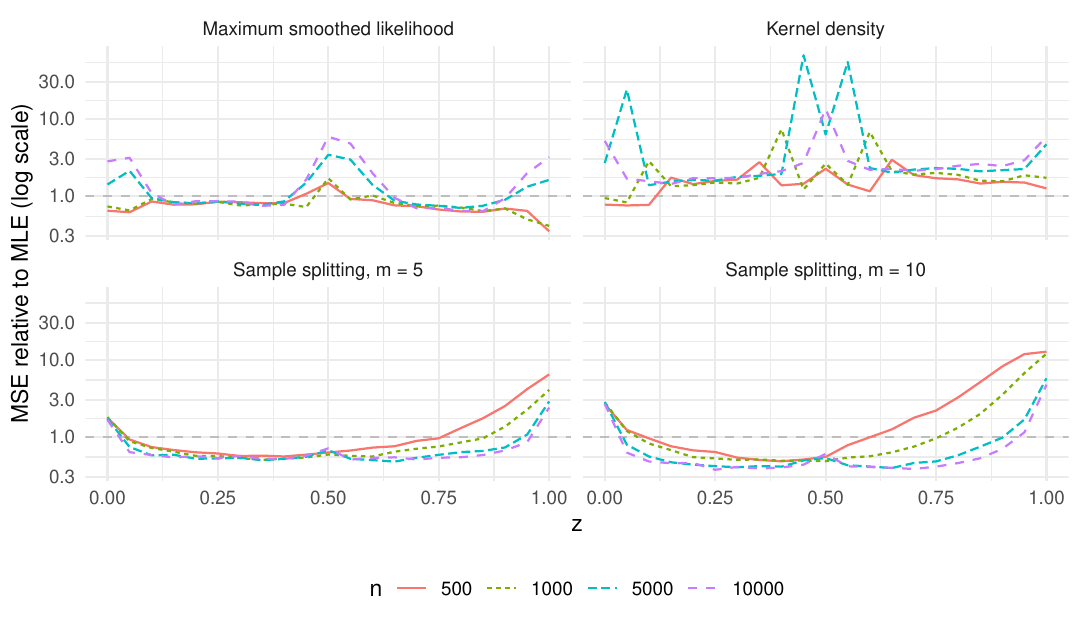}\par
\caption{Relative mean squared errors of the maximum smoothed likelihood estimator, the kernel density-based estimator, and the sample splitting estimators to the maximum likelihood estimator for $z \in [0, 1]$ and various sample sizes $n$. The maximum likelihood has better mean squared error for $y$-values greater than one, and the other estimator has better mean squared error for $y$-values less than one.}
\label{fig:relative_mses}
\end{figure}

Figure~\ref{fig:relative_mses} shows the ratio of the mean squared errors of the maximum smoothed likelihood estimator,  the kernel density-based estimator, and the sample splitting estimators to that of the maximum likelihood estimator. The maximum smoothed likelihood estimator is slightly more efficient than the maximum likelihood estimator at continuity points, but is less efficient around mass points. Furthermore, the relative performance of the maximum likelihood estimator at positive mass points increases as the sample size grows. The kernel density estimator is generally less efficient than the maximum likelihood estimator, especially near mass points, and the discrepancy also grows with the sample size.

For large enough $n$, the sample splitting estimator is more efficient than the maximum likelihood estimator at all points at which the latter is consistent. The relative improvement of $\bar\theta_{n,m}$ grows with the number of splits $m$, as does the sample size $n$ required for $\bar\theta_{n,m}$ to outperform $\theta_n^*$.

\begin{figure}[h!]
\centering
\includegraphics[width=\linewidth]{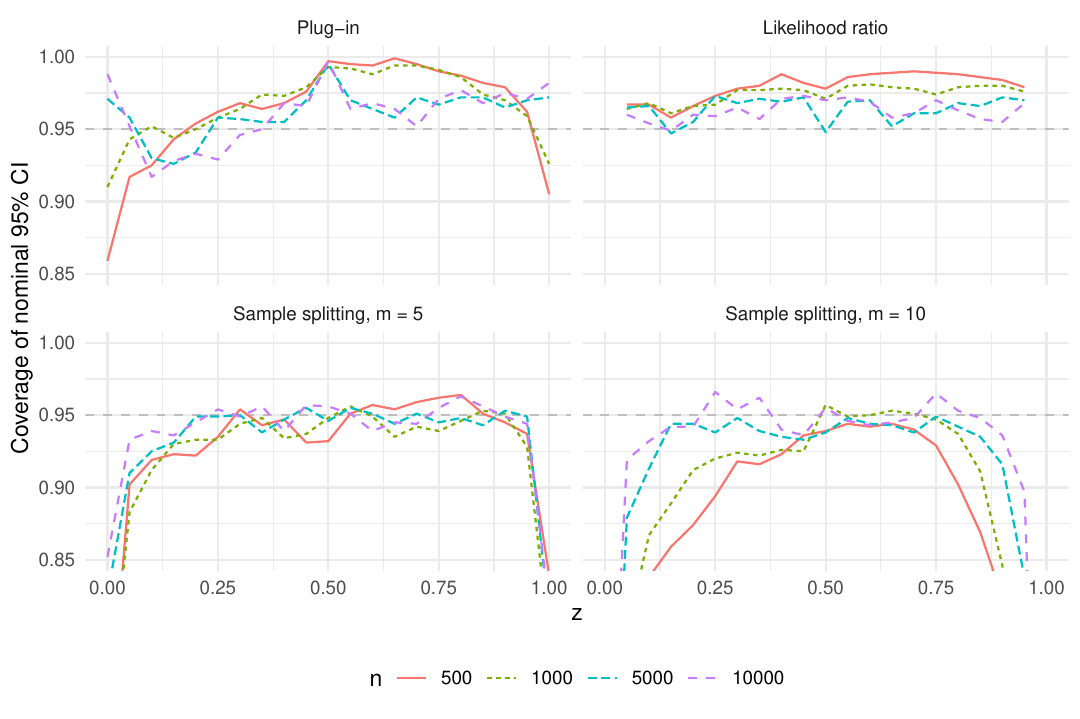}\par
\caption{Coverage of 95\% CIs for $z \in [0, 1]$, various sample sizes $n$, and four methods: the plug-in method centered around the maximum likelihood estimator (upper left), the inverted likelihood ratio tests (upper right), and the sample splitting method with $m=5$ (lower left) and $m=10$ (lower right). Note that the likelihood ratio method does not provide intervals at the endpoints.}
\label{fig:coverages}
\end{figure}

Figure~\ref{fig:coverages} shows the empirical coverage of 95\% confidence intervals for $\theta_0(z)$ constructed using the plug-in method described in Section~\ref{sec:asy_cont}, the inverted likelihood ratio test approach of \cite{banerjee2001ratio}, and the sample splitting approach of \cite{banerjee2019divide} described above. We note that the likelihood ratio approach does not provide intervals at the end points $z = 0$ or $z = 1$. The plug-in method is conservative in large samples near mass points, but anti-conservative at some points of positive density. This is because the plug-in method is designed to work when the distributions are fully continuous, and estimation of the required nuisance parameters in the limit distribution fails in the presence of mass points. The likelihood ratio method is conservative in smaller samples, but approaches nominal coverage in large samples for points $z$ of absolute continuity.  The sample splitting method with $m=5$ has adequate coverage for all sample sizes except for $z$ close to the boundaries. The sample splitting method with $m=10$ (and similarly for $m=20$, which is not shown) appears to require very large sample sizes to attain adequate coverage over a large range of $z$. We note that the sample splitting methods was able to achieve good coverage in large samples at both interior absolutely continuous points and interior mass points, without the user specifying which points are which.

\section{Analysis of C-reactive protein for predicting bacterial infection}\label{sec:analysis}

In this section, we use the methods presented herein to assess the use of the biomarker C-reactive protein (CRP) for determining the presence or absence of bacterial infection in children with systemic inflammatory response syndrome (SIRS). The Optimizing Antibiotic Strategies in Sepsis (OASIS) II study enrolled a prospective observational cohort of children under the age of nineteen at the pediatric intensive care unit at The Children's Hospital of Philadelphia from August 2012 to June 2016 \citep{downes2018oasis}. Patients were enrolled in the  study if they presented signs of SIRS, were started on a new broad-spectrum antibiotic for suspected bacterial infection, and had blood cultures taken within six hours of SIRS onset. A primary goal of the study was to assess whether CRP, which had previously been found to be predictive of bacterial infection \citep{downes2017pragmatic}, could be used to determine when antibiotic therapy could be safely ended. Additional details of the study design and results of the primary analysis may be found in \cite{downes2018oasis}.

We analyzed all patients in the OASIS II cohort with measured biomarkers and bacterial infection status to assess the odds of bacterial infection as a function of CRP value. Some patients had measurements at multiple episodes; since all such episodes were at least 30 days apart, we treated these episodes as independent of one another. We analyzed a total of $n=504$ CRP measurements among 443 unique patients, with $n_1 = 202$ bacterial infections among 191 unique patients and $n_2  = 302$ non-infections among 266 unique patients. 

Since CRP has previously been found to be predictive of bacterial infection in this patient population, there is scientific reason to believe that the density ratio order holds. We therefore computed the MLE of the density ratio function and corresponding 95\% likelihood ratio-based pointwise confidence intervals and the sample splitting estimator of \cite{banerjee2019divide} with $m=5$ splits and corresponding 95\% pointwise confidence intervals.

\begin{figure}[h!]
\centering
\includegraphics[width=\linewidth]{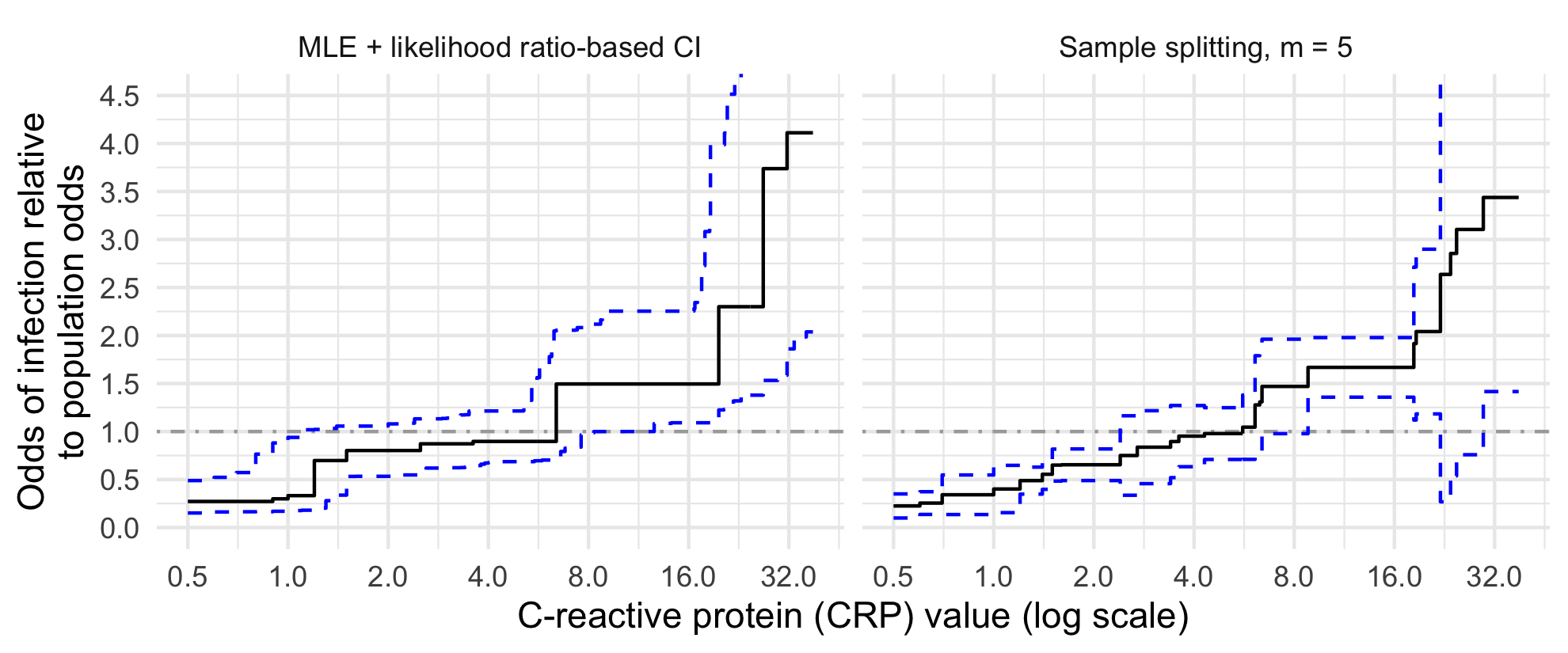}
\caption{Odds of bacterial infection given C-reactive protein value relative to population odds in children with treating systemic inflammatory response syndrome.}
\label{fig:crp}
\end{figure}

Figure~\ref{fig:crp} displays the estimated odds of bacterial infection given CRP value relative to the population odds of bacterial infection and 95\% pointwise confidence intervals. We find that values of CRP under 1 are indicative of roughly quartered odds of infection relative to the population odds of infection, and values of CRP greater than 20 are indicative of roughly doubled odds of infection relative to the population 
odds. Values of CRP between 1 and 20 do not clearly indicate that a patient's odds of infection are larger or smaller than the population odds. 

\section{Discussion}

In this article, we have considered nonparametric maximum likelihood inference for the density ratio function and the individual distribution functions under the assumption that the density ratio is nondecreasing. We applied these methods to the analysis of the biomarker C-reactive protein for predicting bacterial infection in children with systemic inflammatory response syndrome. The methods apply broadly to biomarker analysis, as well as other areas of biomedical research.

One of our important contributions is the ability to deal with discrete, continuous, and mixed discrete-continuous distributions. Such distributions arise frequently in applied settings, and in particular in the context of biomarker analysis. Furthermore, we have demonstrated via numerical studies that sample splitting provides good pointwise confidence interval coverage without knowing which values correspond to discrete mass points and which correspond to points of Lebesgue continuity of the underlying densities, which is important because in practice analysts may not have such knowledge a priori. However, a theoretical treatment of the precise asymptotic behavior of the estimator at mass points remains unknown, and would be an interesting topic of future research.

\singlespacing
{
\section*{Acknowledgments}
The authors thank Craig Boge for help compiling the OASIS II data. The authors also gratefully acknowledge the constructive comments of the editors and anonymous reviewers and the support of the CDC Epicenters program (KJD), NICHD grant K23HD091365 (KJD), the Center for Pediatric Clinical Effectiveness and the Pediatric IDEAS Research Group of the Children's Hospital of Philadelphia (KJD, TW), and the Department of Pediatrics of the University of Pennsylvania Perelman School of Medicine.
\bibliographystyle{apa}
\bibliography{monotone_density_bib}
}

\clearpage

\doublespacing

\begin{center}
\huge{\textsc{Supplementary Material}}
\end{center}

\section*{Example of the use of Theorem~\ref{thm:mle}}

\begin{figure}[h!]
\centering
\includegraphics[width=.49\linewidth]{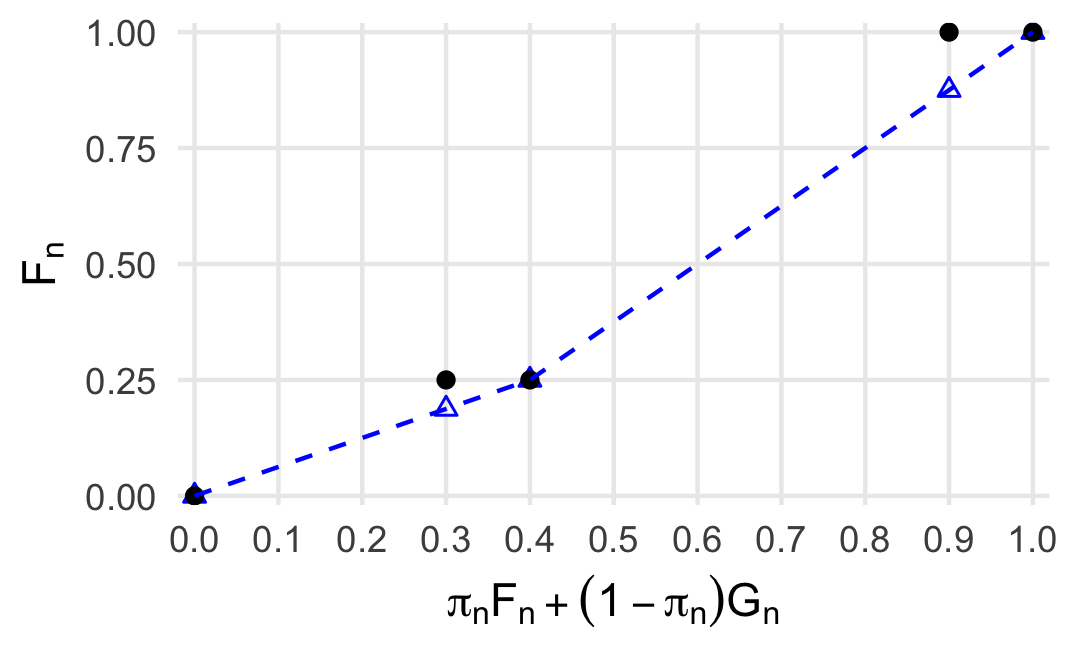}
\includegraphics[width=.49\linewidth]{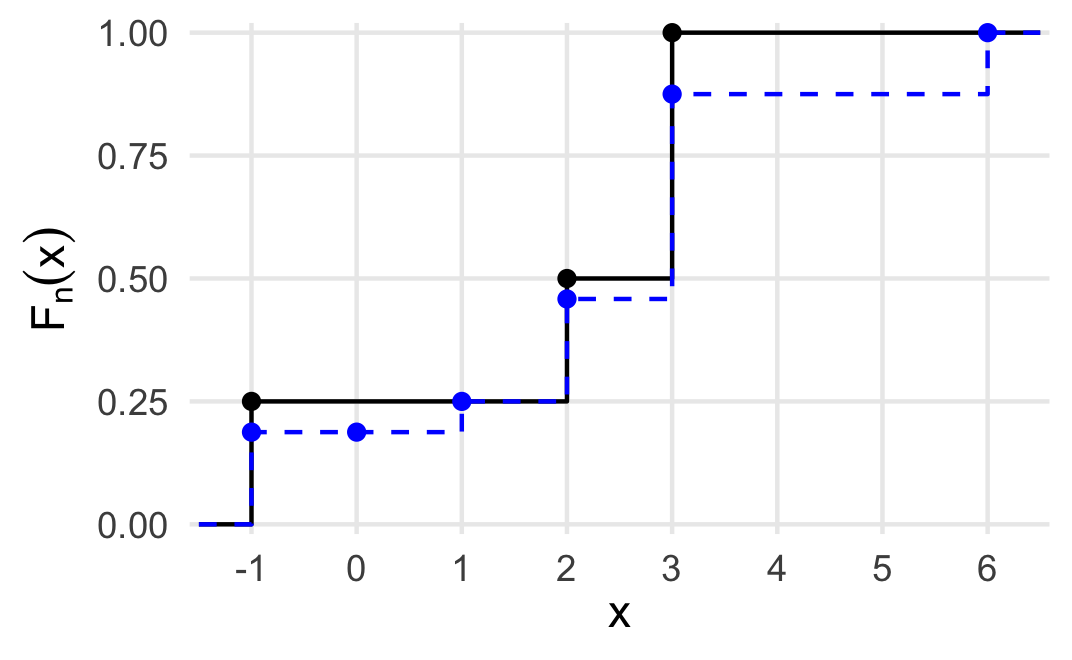}
\includegraphics[width=.49\linewidth]{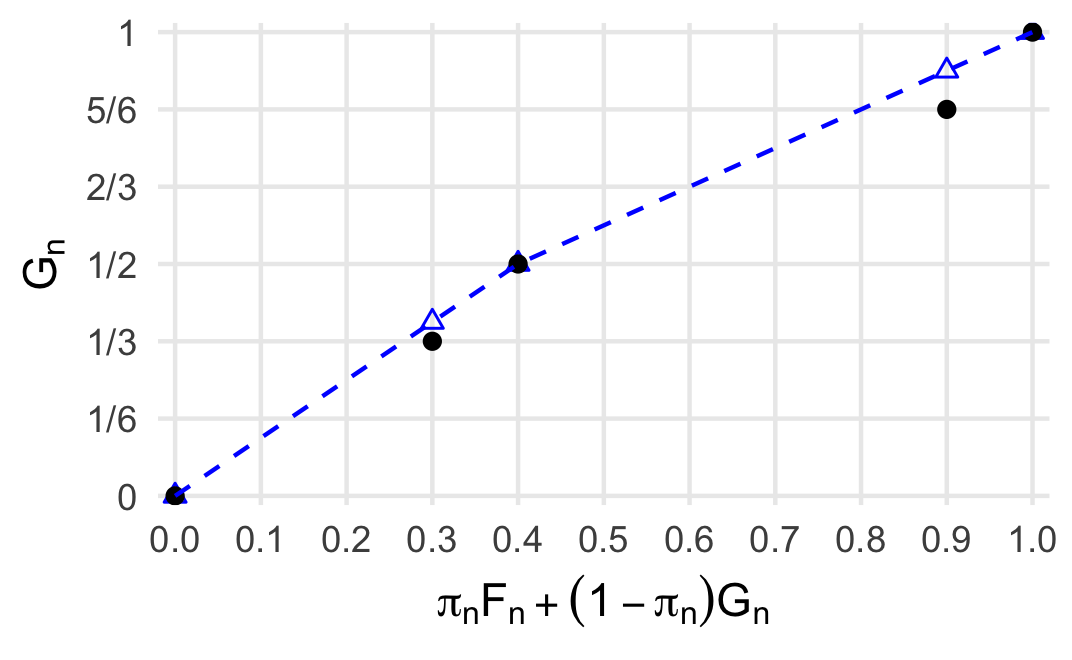}
\includegraphics[width=.49\linewidth]{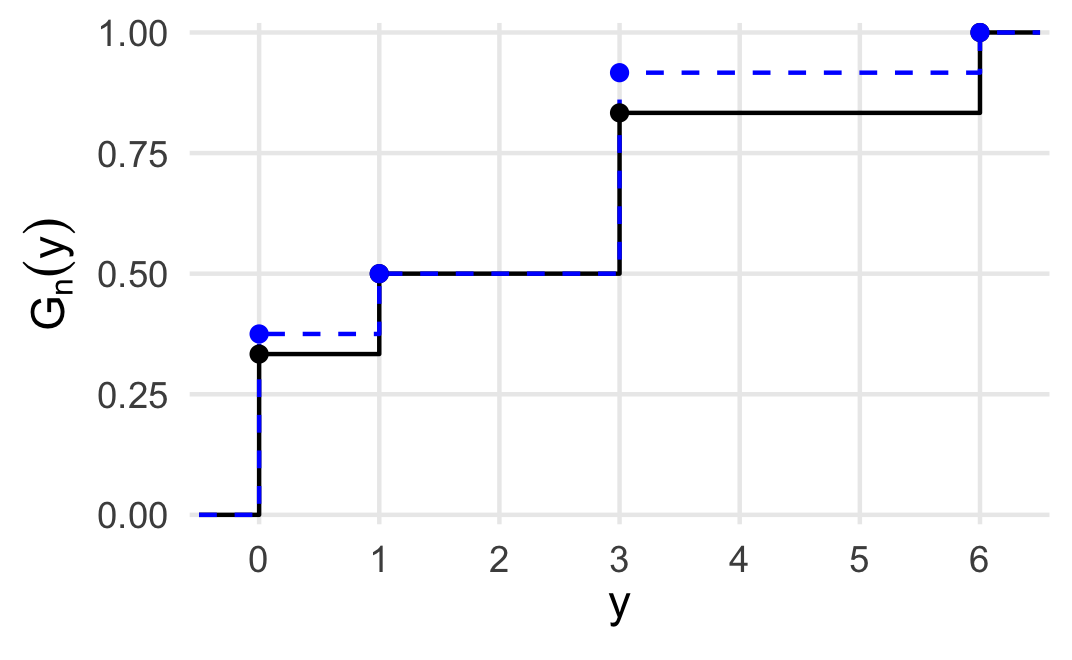}
\includegraphics[width=.49\linewidth]{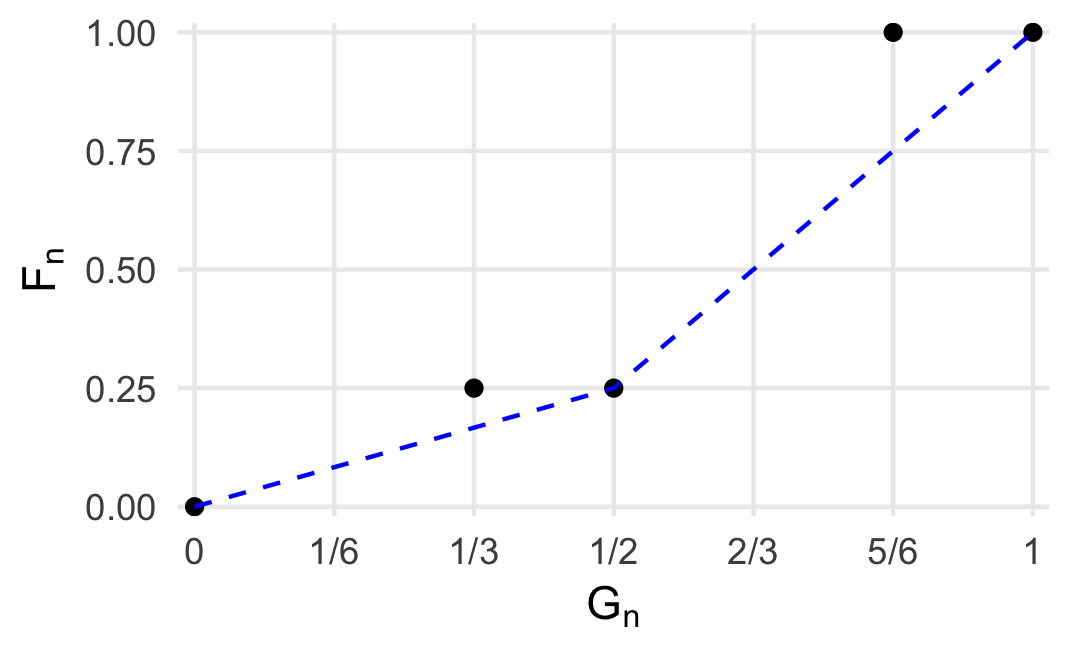}
\includegraphics[width=.49\linewidth]{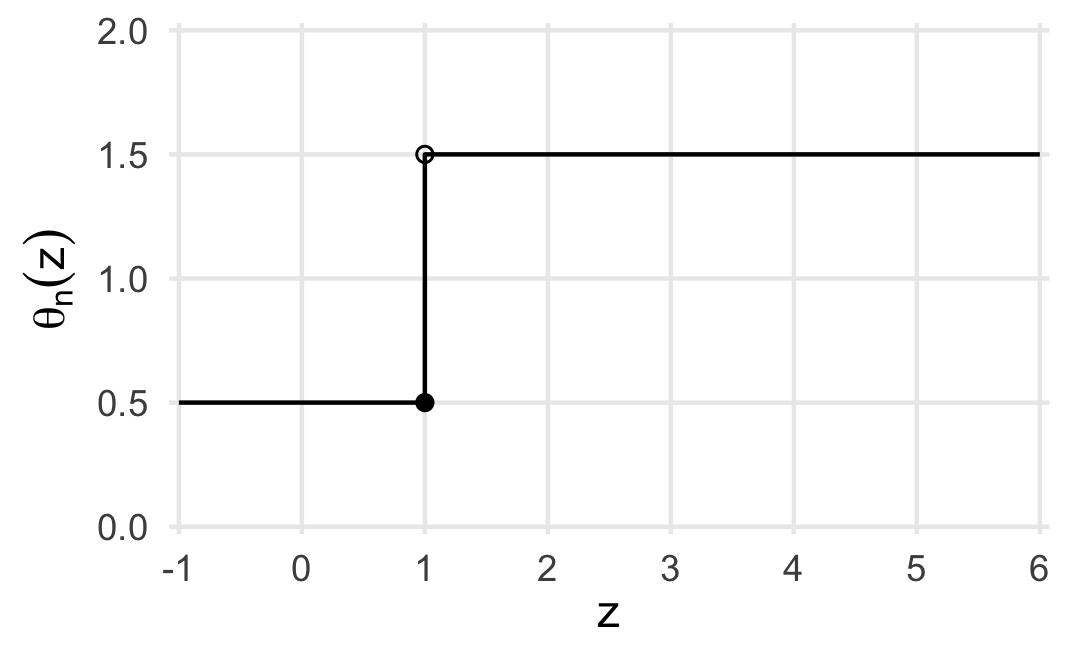}
\caption{Example of the process of constructing the maximum likelihood estimator for $(Y_1, \dots, Y_6) = (0,0,1,3,3,6)$ and $(X_1, \dotsc, X_4) = (-1,2,3,3)$. The graph of $F_n$ versus $\pi_n F_n + (1-\pi_n) G_n$ evaluated at $z_1, \dotsc, z_m$ and its GCM are shown in the upper left. The resulting MLE $F_n^*$ and $F_n$ are shown in the upper right, and the graph of $G_n$ versus $\pi_n F_n + (1-\pi_n) G_n$ evaluated at $z_1, \dotsc, z_m$ and its  LCM are shown in the center left, and the resulting MLE $G_n^*$ and $G_n$ are shown in the middle right. The ODC diagram of $F_n$ versus $G_n$ and its GCM are shown in the bottom left, and the resulting MLE $\theta_n^*$ is shown in the bottom right.}
\label{fig:example}
\end{figure}

We first illustrate the use of Theorem~\ref{thm:mle} and Corollary~\ref{cor:theta_mle} using hypothetical data. Suppose that $(Y_1, \dots, Y_6) = (0,0,1,3,3,6)$ and $(X_1, \dotsc, X_4) = (-1,2,3,3)$. We first derive $F_n^*$. The points $\{ (H_n(y_k), F_n(y_k)) :  k=0, \dotsc, m_2\}$ are given by $\{(0,0),$ $(0.3, 0.25),$ $(0.4, 0.25),$ $(0.9, 1),$ $(1,1)\}$, and its GCM is given by $\{(0,0),$ $(0.3, 3/16),$ $(0.4, 1/4),$ $(0.9, 7/8),$ $(1,1)\}$. This is displayed in the upper left panel of Figure~\ref{fig:example}. The values of the GCM imply that $F_n^*(0) = 3/16$, $F_n^*(1)=1/4$, $F_n^*(3) = 7/8$, and $F_n^*(6) = 1$. We then have that $F_n^*(-1) = F_n^*(-\infty) + [F_n^*(0) - F_n^*(-\infty)]\frac{F_n(-1)- F_n(-1-)}{F_n(0) - F_n(-\infty)} = [3/16]\frac{1/4}{1/4}=3/16$ and $F_n^*(2) = F_n^*(1) + [F_n^*(3) - F_n^*(1)]\frac{F_n(2) - F_n(2-)}{F_n(3) - F_n(1)} = 1/4 + [5/8]\frac{1/4}{3/4} =  11/24$.  The estimators $F_n$ and $F_n^*$ are compared in the bottom left panel of Figure~\ref{fig:example}.

We next derive $G_n^*$. The points $\{ (H_n(y_k), G_n(y_k)) :  k=0, \dotsc, m_2\}$ are given by $\{(0,0),$ $(0.3, 1/3),$ $(0.4, 1/2),$ $(0.9, 5/6),$ $(1,1)\}$, and its LCM is given by $\{(0,0),$ $(0.3, 3/8),$ $(0.4, 1/2),$ $(0.9, 11/12),$ $(1,1)\}$. This is displayed in the center left panel of Figure~\ref{fig:example}. The values of the LCM imply that $G_n^*(0) = 3/8$, $G_n^*(1)=1/2$, $G_n^*(3) = 11/12$, and $G_n^*(6) = 1$. The estimators $G_n$ and $G_n^*$ are compared in the bottom left panel of Figure~\ref{fig:example}.

Finally, we derive $\theta_n^*$. The empirical ordinal dominance curve is given by the points $\{(0,0)$, $(1/3,1/4)$, $(1/2, 1/4)$, $(5/6, 1)$, $(1,1)\}$, and the vertices of its GCM are given by $\{(0,0,)$, $(1/2,1/4)$, $(1,1)\}$. This is displayed in the bottom left panel of Figure~\ref{fig:example}. The left-hand slopes of the GCM are $1/2$ on the interval $(0, 1/2]$ and $3/2$ on the interval $(1/2, 1]$, which implies that $\theta_n^*(z) = 1/2$ for $z \in (-\infty, 1]$ and $\theta_n^*(z) = 3/2$ for $z \in (1, \infty)$.  This is displayed in the bottom right panel of Figure~\ref{fig:example}. 

We note that the maximum likelihood estimators $\hat{F}_n$ of $F_0$ and $\hat{G}_n$ of $G_0$ derived in \cite{dykstra1995mle} for the fully discrete case are different than $F_n^*$ and $G_n^*$. In particular, both $\hat{F}_n$ and $\hat{G}_n$ have jumps at all the unique values of the data $\{-1, 0, 1, 2,3, 6\}$ with values $\hat{F}_n(-1) = 1/16$, $\hat{F}_n(0) = 3/16$, $\hat{F}_n(1) = 1/4$, $\hat{F}_n(2) = 3/8$, and $\hat{F}_n(3) = 7/8$, and $\hat{F}_n(6) = 1$; and $\hat{G}_n(-1) = 1/8$, $\hat{G}_n(0) = 3/8$, $\hat{G}_n(1) = 1/2$, $\hat{G}_n(2) = 7/12$, $\hat{G}_n(3) = 11/12$, and $\hat{G}_n(6) = 1$. However, the maximum likelihood estimator $\hat\theta_n(z) = \Delta \hat{F}_n(z) /  \Delta \hat{G}_n(z)$ is equal to $\theta_n^*(z)$ for each $z \in \{-1, 0, 1, 2,3, 6\}$.

\clearpage

\section*{Proof of Theorems}

\begin{proof}[\bfseries{Proof of Theorem~\ref{thm:lik_ratio_order}}]
We first suppose that $F \ll G$ and $\nu$ is non-decreasing on $\s{G}$, and we show that $F(A) G(B) \leq F(B) G(A)$ for all measurable $A \leq B$. Recall that $F(A) = \int_A \, dF$ when $A$ is a set, and $A \leq B$ means that $a \leq b$ for all $a \in A$ and $b \in B$. Since $F \ll G$, we have that $F(x) = \int_{-\infty}^x \nu(u) \, dG(u)$ for all $x$. We then have by Fubini's Theorem that
\begin{align*}
 F(A) G(B) &= \int_{x \in A} \, dF(x) \int_{y \in B} \, dG(y) = \int_{x \in A} \nu(x) \, dG(x) \int_{y \in B} \, dG(y) \\
 &= \int_{(x,y) \in A \times B} \nu(x) \, d(G \times G)(x,y).
 \end{align*}
Now since $\nu$ is non-decreasing and $x \leq y$ for all $x \in A$ and $y \in B$, we have
\[\int_{(x,y) \in A \times B} \nu(x) \, d(G \times G)(x,y) \leq \int_{(x,y) \in A \times B} \nu(y) \, d(G \times G)(x,y).\]
Finally, applying Fubini's Theorem again yields
\[ \int_{(x,y) \in A \times B} \nu(y) \, d(G \times G)(x,y) =  \int_{x \in A} \, dG(x) \int_{y \in B} \nu(y) \, dG(y) = G(A) F(B).\]

Next, we suppose that $F(A) G(B) \leq F(B) G(A)$ for all measurable $A \leq B$, and we show that $R_{F,G}$ is convex on $\n{Im}(G)$.  Let $t, u, v \in \n{Im}(G)$, where $t < v$ and $u = \lambda t + (1-\lambda) v$ for $\lambda \in (0,1)$. We then let $A = (G^-(t), G^-(u)]$ and $B = (G^-(u), G^-(v)]$, which are both Borel sets satisfying $A \leq B$ since $G^-$ is necessarily non-decreasing. We then have $F(A) = F(G^-(u)) - F(G^-(t)) = R_{F, G}(u) - R_{F,G}(t)$ and similarly $F(B) = R_{F,G}(v) - R_{F,G}(t)$. In addition, since $G(G^-(z)) = z$ for any $z\in \n{Im}(G)$, we also have $G(A) =  G(G^-(u)) - G(G^-(t)) = u - t = (1-\lambda)(v - t)$ and similarly $G(B) = v - u = \lambda (v -t)$. We then have by assumption that
\[ [ R_{F, G}(u) - R_{F,G}(t)] [   \lambda (v -t)] = F(A) G(B) \leq F(B) G(A) =  [ (1-\lambda)(v - t)][ R_{F,G}(v) - R_{F,G}(t)].\]
Therefore, $\lambda \left[ R_{F,G}(u) - R_{F, G}(t) \right] \leq (1-\lambda)\left[ R_{F,G}(v) - R_{F, G}(u) \right]$, which implies that $R_{F, G}(u) \leq \lambda R_{F, G}(t) + (1-\lambda) R_{F, G}(v)$, which shows that $R_{F,G}$ is convex on $\n{Im}(G)$.

Finally, we suppose that $F \ll G$, $R := R_{F, G}$ is convex on $\n{Im}(G)$, and $\nu$ is continuous on $\s{G}$, and we show that $\nu$ is nondecreasing on $\s{G}$. This is the most difficult of the three implications. The basic argument amounts to using convexity of $R$ to compare the slopes of chords or sequences of chords, and to relate these slopes to values of $\nu$. Let $x, y \in \s{G}$ with $x < y$.  Suppose that we can find sequences $\{z_j\}_{j \geq 1}$ and $\{w_j\}_{j\geq 1}$ such that $s_j := [R(G(x)) - R(G(z_j))] / [G(x) - G(z_j)]$ converges to $\nu(x)$, $t_j := [R(G(y)) - R(G(w_j))] / [G(y) - G(w_j)]$ converges to $\nu(y)$, and $z_j \leq w_j$ for all $j$ large enough. Then, by convexity of $R$, $s_j \leq t_j$ for all $j$ large enough, which implies that $\nu(x) \leq \nu(y)$. The exact form of $\{z_j\}_{j \geq 1}$ and $\{w_j\}_{j\geq 1}$ depends on how $G$ looks near $x$ and $y$. In particular, there are three cases for $y$: (1) $G(y) > G(y-)$ and there exists $p \in [x, y)$ such that $G(y-) = G(p)$; (2) $G(y) > G(y-)$ but there is no $p \in [x,y)$ such that $G(y-) = G(p)$; and (3) $G(y) = G(y-)$. We begin by specifying $\{w_j\}_{j\geq 1}$ in each case.

In case (1), we take $w_j = p$ for all $j$. Since $F \ll G$, we must have $F(G^-(G(p))) = F(y-)$, so that $t_j = \nu(y)$ for all $j$.  In case (2), it must be that $G^-(G(y-)) = y$. In this case, there exists $\{w_j\}_{j\geq 1}$ increasing to $y$ such that $w_j \in (x,y) \cap \s{G}$ for each $j$, $G(w_j)$ increases to $G(y-)$ and $F(w_j)$ increases to $F(y-)$. We then have that $R(G(w_j))$ increases to $F(G^-(G(y-))-) = F(y-)$, so that $t_j$ increases to $[F(y) - F(y-)]/ [G(y) - G(y-)] = \nu(y)$. In case (3), we first note that $F(G^-(G(y))) = F(y)$ since $F \ll G$. Additionally, since $y \in\s{G}$, there exist $\{w_j\}_{j \geq 1}$ in $\s{G}$ with $G^-(G(w_j)) = w_j$ for each $j$ that either (a) increases to $y$ and $G(w_j) < G(y)$ for each $j$, or (b) decreases to $y$ and $G(w_j) > G(y)$ for each $j$. In either case, we have
\[ t_j = \frac{\int_{w_j}^{y} \nu(u) \, dG(u)}{G(y) - G(w_j)} = \nu(y) +\frac{\int_{w_j}^{y} [\nu(u)-\nu(y)] \, dG(u)}{G(y) - G(w_j)} \ . \]
For any $\varepsilon > 0$, by continuity of $\nu$ over $\s{G}$, we can find $m$ such that $j \geq m$ implies $|\nu(u) - \nu(y)| < \varepsilon$ for all $u \in [w_j, y] \cap\s{G}$. If (a) holds and $t_j$ is bounded above, we then have $\int_{w_j}^{y} |\nu(u)-\nu(y)| \, dG(u) \leq \varepsilon [G(y) - G(w_j)]$ for all $j \geq m$, so that then $\lim_{j \to \infty} t_j = \nu(y)$. If $t_j$ is not bounded above then $\nu(y) = +\infty$, so that $\nu(x) \leq \nu(y)$ trivially. If (b) holds then $t_j$ is bounded below by zero, so by a similar calculation $\lim_{j \to \infty} t_j = \nu(y)$.  

The three cases for $x$ are similar: (1) $G(x) > G(x-)$ and there exists $q \in  [-\infty, x)$ such that $G(x-) = G(q)$; (2) $G(x) > G(x-)$ but there is no such $q$; and (3) $G(x) = G(x-)$. In case (1), we take $z_j = q$ for all $j$. Since $F \ll G$, we must have $F(G^-(G(q))) = F(x-)$, so that $s_j = \nu(y)$ for all $j$. In case (2), it must be that $G^-(G(x-)) = x$, and again there exists an increasing sequence $\{z_j\}_{j\geq1}$ increasing to $x$ such that $z_j \in (-\infty, x) \cap \s{G}$ for each $j$, $G(z_j)$ increases to $G(x-)$ and $F(z_j)$ increases to $F(x-)$. We then have that $R(G(z_j))$ increases to $F(x-)$, so that $s_j$ increases to $\nu(x)$. In case (3), $F(G^-(G(x))) = F(x)$, and since $x \in\s{G}$, there exists $\{z_j\}_{j\geq 1}$ in $\s{G}$ with $G^-(G(z_j)) = z_j$ for each $j$ that either (a) increases to $x$ and $G(z_j) < G(x)$ for each $j$, or (b) decreases to $x$ and $G(z_j) > G(x)$ for each $j$. If (a) holds and $s_j$ is bounded above, then $s_j$ converges to $\nu(x)$ by continuity of $\nu$ as before. If $s_j$ is not bounded above then $s_j$ converges to $\nu(x) = +\infty$. If (b) holds then $s_j$ is bounded below by zero, so again $\lim_{j \to \infty} s_j = \nu(x)$.

Of the nine pairings of cases for $y$ and cases for $x$, the only situation in which it is not immediately clear that $z_j \leq w_j$ for all $j$ large enough is that $z_j$ decreases to $x$ (case 3b) and $w_j = p$ for all $j$ (case 1). However, we note that $x = p$ if and only if $G(x) = G(y-)$, which would imply that case (3b) cannot hold for $x$. Therefore, if $z_j$ decreases to $x$ and $w_j = p$, then $p > x$, so that $z_j < w_j$ for all $j$ large enough. This completes the argument.

Finally, we address statement (2) of the result: we suppose that $F \ll G$ and $\nu$ is continuous and non-decreasing on $\s{G}$, and we show that $\theta_{F,G} = \nu$ on $\s{G}$. By (1), $R$ is convex on $\n{Im}(G)$. First, we claim that $\n{GCM}_{[0,1]}(R) = H$, where $H: [0,1] \to [0,1]$ takes the following form. For any $ u \in \n{Im}(G)$, $H(u) := R(u)$. If $u \notin \n{Im}(G)$, then there exists $x \in \d{R}$ and $\lambda \in [0,1)$ such that $u = \lambda G(x-) + (1-\lambda) G(x)$. We then define $H(u) := \lambda R(G(x-)-) +(1-\lambda) R(G(x))$. Thus, $H$ is the linear interpolation of $R|_{\n{Im}(G)}$ to $[0,1]$. In order to show that $H$ indeed equals $\n{GCM}_{[0,1]}(R)$, we need to show that (a) $H$ is  convex, (b) $H \leq R$, and (c) $H \geq \bar{H}$ for any other convex minorant of $R$.

For (a), we let $u, v \in [0,1]$ and $p = \lambda u + (1-\lambda) v$ for $\lambda \in (0,1)$. There then exist $u_1 \leq u_2 \leq p_1 \leq p_2 \leq v_1 \leq v_2$ which are all elements of $\n{Im}(G)$ and $\lambda_1, \lambda_2, \lambda_3 \in [0,1]$ such that $u = \lambda_1u_1 + (1-\lambda_1)u_2$, $v = \lambda_2 v_1 + (1-\lambda_2) v_2$, and $p = \lambda_3 p_1 + (1-\lambda_3)p_2$, and furthermore  $H(u) = \lambda_1R(u_1-) + (1-\lambda_1)R(u_2)$, $H(v) = \lambda_2 R(v_1-) + (1-\lambda_2) R(v_2)$, and $H(p) = \lambda_3 R(p_1-)+ (1-\lambda_3)R(p_2)$. The remainder of the argument is best seen with a diagram. Let $U$ be the point $(u, H(u))$, $U_1$ be the point $(u_1, H(u_1))$, and so on. By convexity of $R$, the line segment $\overline{P_1P_2}$ lies below or on the line segment $\overline{U_2V_1}$, which lies below or on $\overline{UV_1}$, which lies below or on $\overline{UV}$. Therefore, $(p, H(p))$, which falls on $\overline{P_1P_2}$, is no greater than $(p, \lambda H(u) + (1-\lambda) H(p))$, which falls on $\overline{UV}$.

For (b), by definition, $H(u) = R(u)$ for any $u \in \n{Im}(G)$. If $u \notin \n{Im}(G)$, then $u = \lambda G(x-) + (1-\lambda) G(x)$, and  hence  $G^-(u) = G^-(G(x)) = x$. As a result, $R(u) = R(G(x)) > H(u) = \lambda R(G(x-)-) +(1-\lambda) R(G(x))$.

We have now shown that $H$ is a convex minorant of $R$. For (c), if $\bar{H}$ is another convex minorant of $R$, then clearly $H(u) \geq \bar{H}(u)$ for all $u \in \n{Im}(G)$. If $u \notin \n{Im}(G)$, then $u = \lambda G(x-) + (1-\lambda) G(x)$. If $G(x-) \in \n{Im}(G)$, then $\bar{H}(u) \leq \lambda H(G(x-)) + (1-\lambda) H(G(x)) \leq \lambda R(G(x-)) + (1-\lambda) R(G(x)) = H(u)$. If $G(x-) \notin \n{Im}(G)$, then there must be an $\varepsilon > 0$ such that $z \in \n{Im}(G)$ for all $z \in (G(x-) - \varepsilon, G(x-))$, so that $\bar{H}(u) \leq \lambda(z) R(z-) + (1-\lambda(z)) R(G(x))$ for each $z \in (G(x-) - \varepsilon, G(x-))$, where $\lambda(z) \in (0,1)$ and $\lambda(z) \to \lambda$ as $z \to G(x-)$. Taking the limit as $z \to G(x-)$, we have that  $\bar{H}(u) \leq \lambda R(G(x-)-) + (1-\lambda) R(G(x)) = H(u)$.

We now have that $\theta_{F,G}(x) = (\partial_-H)(G(x))$, so it remains to show that $(\partial_-H)(G(x)) = \nu(x)$ for all $x \in \s{G}$. First, if $G(x) > G(x-)$, then $H(u) = \lambda R(G(x-)-) + (1-\lambda) R(G(x)) = \lambda F(x-) + (1-\lambda) F(x)$ for all $u = \lambda G(x-) + (1-\lambda)G(x)$ for $\lambda \in (0,1)$. Therefore, $(\partial_- H)(u) = [F(x) - F(x-)] / [G(x) - G(x-)] = \nu(x)$ for all such $u$, so that $(\partial_- H)(G(x)) = \nu(x)$. If instead $x \in \s{G}$ and $G(x) = G(x-)$ then $H(G(x)) = R(G(x)),$ and it is straightforward to see from the definition of $R$ that $(\partial_-R)(G(x)) = \nu(x)$.
\end{proof}

\begin{proof}[\bfseries{Proof of Theorem~\ref{thm:mle}}]
We first note that $L_n(F, G) = 0$ for any $G$ such that $G(Y_j) = G(Y_j-)$ for any $j \in \{1, \dotsc, n_2\}$. As a result, we may restrict our attention to $G$ such that $G(Y_j) > G(Y_j-)$ for all $j$, which implies that $G^-$ has support at each $G(Y_j)$. For any such $G$, we define $\bar{G} := G \circ L$, where $L(y) := \max\{ Y_j : Y_j \leq y\}$. We then have $\bar{G}(Y_j) - \bar{G}(Y_j- ) \geq G(Y_j) - G(Y_j-)$ for each $j$. Furthermore, the support of $\bar{G}^-$ is $\{G(Y_j) : j = 1, \dotsc, n_2\}$ is contained in the support of $G$, $\bar{G}(Y_j) = G(Y_j)$ for each $j$, and $F\circ G^-$ is by assumption convex on the support of $G^-$. Therefore, $F \circ \bar{G}^-$ is convex on the support of $\bar{G}^-$, so that $(F, \bar{G}) \in \s{M}_0$ and $L_n(F, \bar{G}) \geq L_n(F, G)$. Hence, we may further restrict our attention to $G$ which are discrete with jumps at $Y_1, \dotsc, Y_{n_2}$. By a similar argument, we can restrict our attention to $F$ which are discrete with jumps at $X_1, \dotsc, X_{n_1}$ \emph{or} $Y_1, \dotsc, Y_{n_2}$.

We define $y_0 := -\infty$, and $u_j := G(y_j)$, so that the support of $G^-$ for any discrete $G$ with jumps at $Y_1, \dotsc, Y_{n_2}$ is $\{u_j : j = 0, \dotsc, m_2\}$, and $G^-(u_j) = y_j$. Defining $g_j := u_j - u_{j-1}$ and $s_j$ the number of $Y_k$ such that $Y_k = y_j$, we have $\prod_{j=1}^{n_2} \left[ G(Y_j) - G(Y_j-)  \right] = \prod_{j=1}^{m_2} g_j^{s_j}$. We then define $f_j := F(y_j) - F(y_j-)$  for each $j$, and we note that $(F, G) \in \s{M}_0$ if and only if $f_1/ g_1 \leq f_2 / g_2 \leq \cdots \leq f_{m_2} / g_{m_2}$. Suppose that the values $f_1, \dotsc, f_{m_2}$ are fixed in such a way as to satisfy these constraints. We denote by $\s{I}_j := \{ k : x_k \in (y_{j-1}, y_{j}]\}$ for $j = 1, \dotsc, m_2+1$, where $y_{m_2+1} := +\infty$, and by $r_i$ the number of $X_k$ such that $X_k = x_i$. Noting that $\s{I}_1, \dotsc, \s{I}_{m_2+1}$ are disjoint with union $\{1, \dotsc, m_1\}$,  we then have 
\[ \prod_{i=1}^{n_1}  \left[ F(X_i) - F(X_i-) \right] = \prod_{j=1}^{m_2+1} \prod_{k \in \s{I}_j}  \left[ F(x_k) - F(x_k-) \right]^{r_k} \ .\]
Additionally, for each $j \in \{1,\dotsc, m_2+1\}$, we must have that $\sum_{k \in \s{I}_j} \left[ F(x_k) - F(x_{k}-) \right]= f_{j}$. Therefore, maximizing $L_n(F, G)$ with respect to $F$ with $f_1, \dotsc, f_{m_2+1}$ fixed amounts to maximizing $ \prod_{k \in \s{I}_j}  \left[ F(x_k) - F(x_{k}-) \right]^{r_k}$ subject to $\sum_{k \in \s{I}_j} \left[ F(x_k) - F(x_{k}-) \right]= f_{j}$ for each $j$. This implies that a maximizer $F_n^*$ must satisfy
\[F_n^*(x_k) - F_n^*(x_{k}-) = f_{j} \frac{r_k}{\sum_{l\in\s{I}_j} r_l} \]
 for each $x_k \in \s{I}_j$. Therefore, $ \prod_{k \in \s{I}_j}  \left[F_n^*(x_k) - F_n^*(x_{k}-) \right]^{r_k}$ is proportional to $ \prod_{k \in \s{I}_j}  f_{j}^{r_k} = f_{j}^{R_j}$ for $R_j := \sum_{k \in \s{I}_j} r_k$, which is the number of $X_i$ in the interval $(y_{j-1}, y_{j}]$.
 
 We note that if there are $j$ such that no $x_k \in (y_j, y_{j+1}]$ but $f_j >0$, then there are infinitely many maximizers  because any $F_n^*$ that assigns mass $f_j$ to the interval $(y_{j-1}, y_{j}]$ yields the same likelihood and satisfies the constraints. In these cases, for the sake of uniqueness we will put mass $f_{j}$ at the point $y_{j}$.
 
 We have at this point reduced the problem to maximizing
 \[ \ \left\{ \prod_{k=1}^{m_2+1} f_k ^{R_k}\right\}\left\{ \prod_{k=1}^{m_2} g_k^{s_k}\right\} = \left\{ \prod_{k=1}^{m_2} f_k^{R_k} g_k^{s_k} \right\} f_{m_2 + 1}^{R_{m_2 + 1}}\]
subject to $f_1/ g_1 \leq f_2 / g_2 \leq \cdots \leq f_{m_2} / g_{m_2}$ and $\sum_{k=1}^{m_2} g_k = \sum_{k=1}^{m_2 + 1} f_k = 1$. Letting $\bar{f}_k := f_k / (1-f_{m_2+1})$ for $k \leq m_2$, this is equivalent to maximizing 
 \[ \bar{L}_n(\bar{f}_1, \dotsc, \bar{f}_{m_2}, f_{m_2 + 1}, g_1, \dotsc, g_{m_2}) := \left\{ \prod_{k=1}^{m_2} \bar{f}_k^{R_k} g_k^{s_k} \right\} (1-f_{m_2 + 1})^{n_1 - R_{m_2 + 1}}f_{m_2 + 1}^{R_{m_2 + 1}}\]
subject to $\bar{f}_1/ g_1 \leq \bar{f}_2 / g_2 \leq \cdots \leq \bar{f}_{m_2} / g_{m_2}$ and $\sum_{k=1}^{m_2} g_k = \sum_{k=1}^{m_2} \bar{f}_k = 1$. The term involving $f_{m_2+1}$ is maximized for $f_{m_2+1}^* = R_{m_2 + 1} / n_1 = 1 - F_n(y_{m_2})$.

From this point we take a similar approach to that in \cite{dykstra1995mle}. We define $\bar{n}_1 := \sum_{k=1}^{m_2} R_k = F_n(y_{m_2}) n_1$, $\sigma_k := \bar{n}_1 \bar{f}_k + n_2 g_k$ and $\rho_k := \bar{n}_1 \bar{f}_k / \sigma_k$, so that $\bar{f}_k = \rho_k \sigma_k / \bar{n}_1$ and $g_k = (1-\rho_k) \sigma_k / n_2$. Optimizing $\bar{L}_n$ with with respect to $\bar{f}_1, \dotsc, \bar{f}_{m_2}$ and $g_1, \dotsc, g_{m_2}$ such that $\sum_{k=1}^{m_2} \bar{f}_k = \sum_{k=1}^{m_2} g_k = 1$ and $\bar{f}_1 / g_1 \leq \bar{f}_2 / g_2 \leq \cdots \leq \bar{f}_{m_2} / g_{m_2}$ is equivalent to optimizing 
\[\bar{L}_n(\bs\rho, \bs\sigma) = \prod_{k=1}^{m_2} \left[ \rho_k \sigma_k /  \bar{n}_1\right]^{R_k} \left[ (1-\rho_k) \sigma_k / n_2\right]^{s_k} =  \bar{n}_1^{- \bar{n}_1} n_2^{-n_2} \prod_{k=1}^{m_2} \rho_k^{R_k}(1-\rho_k)^{s_k} \prod_{k=1}^{m_2}  \sigma_k^{R_k + s_k} \] 
such that $\sum_{k=1}^{m_2} \rho_k \sigma_k =  \bar{n}_1$, $\sum_{k=1}^{m_2} \sigma_k =  \bar{n}_1 + n_2$, and $\rho_1  \leq \cdots \leq \rho_{m_2}$, where $\bs\rho := (\rho_1, \dotsc, \rho_{m_2})$ and $\bs\sigma := (\sigma_1, \dotsc, \sigma_{m_2})$.

Now, $\prod_{k=1}^{m_2} \sigma_k^{R_k + s_k}$ such that $\sum_{k=1}^{m_2} \sigma_k =  \bar{n}_1 + n_2$ is maximized for $\sigma_k^* = R_k + s_k$. Next, maximizing $\prod_{i=1}^{m_2} \rho_k^{R_k}(1-\rho_k)^{s_k}$ with respect to $\rho_1 \leq \cdots \leq \rho_{m_2}$ is equivalent to maximizing
\[ \sum_{k=1}^{m_2} \left[ R_k \log \rho_k + s_k \log (1 - \rho_k)\right] = \sum_{k=1}^{m_2} w_k \left[ t_k \log \rho_k + (1-t_k) \log (1 - \rho_k)\right] \]
for $w_k := R_k + s_k \geq 1$ and $t_k := R_k / w_k$. By Theorem~2.1 and Exercise~2.21 of \cite{groene2014shape}, the maximizer $(\rho_1^*, \dotsc, \rho_{m_2}^*)$ of this expression over all $\rho_1 \leq \cdots \leq \rho_{m_2}$ is given by the weighted isotonic regression of $t_1, \dotsc, t_{m_2}$ with weights $w_1, \dotsc, w_{m_2}$. By Lemma~2.1 of \cite{groene2014shape}, $\rho_k^*$ is equal to the left derivative  of the GCM of the set of points
\begin{align*}
&\left\{ (0,0) \right\} \cup \left\{ \left( \sum_{j=1}^k w_k, \sum_{j=1}^k t_j w_j \right) : k=1, \dotsc, m_2 \right\} \\
&\qquad\qquad= \left\{ \left(n_1 F_n(y_k) + n_2 G_n(y_k), n_1 F_n(y_k)\right) : k=0, \dotsc, m_2\right\}
\end{align*}
evaluated at $n_1 F_n(y_k) + n_2 G_n(y_k)$. We note that $\sum_{k=1}^{m_2} w_k \rho_k^* = \sum_{k=1}^{m_2} \sigma_k^* \rho_k^*  = n_1 F(y_{m_2}) = \bar{n}_1$. Therefore, we have that $L_n(\bs\rho, \bs\sigma) \leq L_n( \bs\rho^*, \bs\sigma^*)$ for all $\bs\rho$ such that $\rho_1  \leq \cdots \leq \rho_{m_2}$ and $\bs\sigma$ such that $\sum_{k=1}^{m_2} \sigma_k = \bar{n}_1 + n_2$. Since $\bs\rho^*$ and $\bs\sigma^*$ also satisfy $\sum_{k=1}^{m_2} \sigma_k^* \rho_k^* = \bar{n}_1$, this implies that $(\bs\rho^*, \bs\sigma^*)$ is an optimizer of $\bar{L}_n$ over the set of stated constraints.

We now have that $f_k^* =(R_k + s_k)(\rho_k^*/n_1)$ and $g_k^* = (R_k + s_k)(1-\rho_k^*)/n_2$. Since $w_k = R_k + s_k$, this implies that $F_n^*(y_k) = \bar{A}_k / n_1$ and $G_n^*(y_k) = [n_2G_n(y_k) + n_1 F_n(y_k)   -\bar{A}_k ]/ n_2$, where $\bar{A}_k$ is the value of the GCM of the set of points defined above at $n_1 F_n(y_k) + n_2 G_n(y_k)$. We note that $\bar{A}_k / n_1 = A_k^*$, for $A_k^*$ the value of the GCM of $\left\{ \left(\pi_n F_n(y_k) + (1-\pi_n)G_n(y_k), F_n(y_k)\right) : k=0, \dotsc, m_2\right\}$ evaluated at $\pi_n F_n(y_k) + (1-\pi_n)G_n(y_k)$. Additionally, $[n_2G_n(y_k) + n_1 F_n(y_k)   -\bar{A}_k ]/ n_2 = B_k^*$ for $B_k^*$ the value of the LCM of $\left\{ \left(\pi_n F_n(y_k) + (1-\pi_n)G_n(y_k), G_n(y_k)\right) : k=0, \dotsc, m_2\right\}$ at $\pi_n F_n(y_k) + (1-\pi_n)G_n(y_k)$.
\end{proof}

\begin{proof}[\bfseries Proof of Corollary~\ref{cor:theta_mle}]
From the proof of Theorem~\ref{thm:mle}, we have that $F_n^*(y_k) = A_k^*$ and $G_n^*(y_k) = G_n(y_k) + \frac{\pi_n}{1-\pi_n} [F_n(y_k) - A_k^*]$. Let $j_0', \dotsc, j_K'$ denote the indices of the vertices of the GCM of $\left\{ \left(\pi_n F_n(y_k) + (1-\pi_n)G_n(y_k), F_n(y_k)\right) : k=0, \dotsc, m_2\right\}$. Then $F_n^*(y_{j_k}) = F_n(y_{j_k})$ for each $k = 0, \dotsc, K$ and $G_n^*(y_{j_k}) = G_n(y_{j_k})$. It is also straightforward to see that $\{ (h_k, A_k) : k = 0, \dotsc, m_2\}$ is a convex minorant of $\{(h_k, F_n(y_k)) : k = 0, \dotsc, m_2\}$ if and only if $\{(G_n(y_k), A_k) : k = 0, \dotsc, m_2\}$  is a convex minorant of $\{(G_n(y_k), F_n(y_k)) : k = 0, \dotsc, m_2\}$. Therefore, $\{(F_n(y_{j_k}), G_n(y_{j_k})) : k= 0, \dotsc, K\}$ form the vertices of the GCM of $\left\{ \left(G_n(y_k), F_n(y_k)\right) : k=0, \dotsc, m_2\right\}$.
\end{proof}

\begin{proof}[\bfseries Proof of Theorem~\ref{thm:discrete}]
We note that $G_n(y_j) > 0$ for each $j$ with probability tending to one. Then, since the support $\s{G}$ of $G_0$ is finite, with probability tending to one the empirical ODC is a left-continuous step function with vertices at $(0,0), (G_n(y_1), F_n(y_1)),$ $\dotsc, (G_n(y_{m_2}), F_n(y_{m_2})$, where we note that $G_n(y_{m_2}) = 1$ almost surley. We define 
\[\delta := \min\left\{ \frac{F_0(y_{j+1}) - F_0(y_{j})}{\Delta G_0(y_{j+1})} -  \frac{F_0(y_{j}) - F_0(y_{j-1})}{\Delta G_0(y_{j})} : j = 1, \dotsc, m_2-1\right\},\]
which is positive by assumption. We then have
\begin{align*}
\frac{F_n(y_{j+1}) - F_n(y_{j})}{\Delta G_n(y_{j+1})} -  \frac{F_n(y_{j}) - F_n(y_{j-1})}{\Delta G_n(y_{j})} &= \frac{F_0(y_{j+1}) - F_0(y_{j})}{\Delta G_0(y_{j+1})} -  \frac{F_0(y_{j}) - F_0(y_{j-1})}{\Delta G_0(y_{j})} \\
&\qquad + \frac{[F_n(y_{j+1}) - F_0(y_{j+1})]- [F_n(y_{j}) - F_0(y_{j})]}{\Delta G_0(y_{j+1})} \\
&\qquad + [ F_n(y_{j+1}) - F_n(y_j)] \left[ \frac{1}{\Delta G_n(y_{j+1})} -   \frac{1}{\Delta G_0(y_{j+1})} \right] \\
&\qquad - \frac{[F_n(y_{j}) - F_0(y_{j})]- [F_n(y_{j-1}) - F_0(y_{j-1})]}{\Delta G_0(y_{j})} \\
&\qquad - [ F_n(y_{j}) - F_n(y_{j-1})] \left[ \frac{1}{\Delta G_n(y_{j})} -   \frac{1}{\Delta G_0(y_{j})} \right].
\end{align*}
Now since $F_n$ is uniformly consistent for $F_0$ and $G_n$ is uniformly consistent for $G_0$,  and $\Delta G_0(y_j) > 0$ for each $j$, the second through fifth  lines above are $\fasterthan(1)$ uniformly over $j$. Therefore, 
\[ \min\left\{ \frac{F_n(y_{j+1}) - F_n(y_{j})}{\Delta G_n(y_{j+1})} -  \frac{F_n(y_{j}) - F_n(y_{j-1})}{\Delta G_n(y_{j})} : j = 1, \dotsc, m_2 -1 \right\} \geq \delta  - \fasterthan(1),\]
which implies that
\[ \frac{F_n(y_{j+1}) - F_n(y_{j})}{\Delta G_n(y_{j+1})} \geq \frac{F_n(y_{j}) - F_n(y_{j-1})}{\Delta G_n(y_{j})}\]
for all $j = 1, \dotsc, m_2 - 1$ with probability tending to one. Therefore, with probability tending to one, the diagram of points $(0,0), (G_n(y_1), F_n(y_1))$, $\dotsc, (G_n(y_{m_2}), F_n(y_{m_2}))$ is convex. Now by Corollary~\ref{cor:theta_mle}, the points $(G_n^*(y_k), F_n^*(y_k))$ lie on the GCM of $(0,0), (G_n(y_1), F_n(y_1))$, $\dotsc, (G_n(y_{m_2}), F_n(y_{m_2}))$. But these points being convex means that they are equal to their GCM, so that with probability tending to one $G_n^*(y_j) = G_n(y_j)$ and $F_n^*(y_j) = F_n(y_j)$ for each $j$. We can then see by Theorem~\ref{thm:mle} that $\Delta F_n^*(x_i) = \Delta F_n(x_i)$ with probability tending to one as well, so that $F_n^* = F_n$ with probability tending to one. We then have with probability tending to one that 
\[\theta_n^*(y_j) = \left[\partial_- \n{GCM}_{[0,1]}( R_{F_n, G_n})\right] \circ G_n(y_j) = \frac{F_n(y_j) - F_n(y_{j-1})}{\Delta G_n(y_j)}\]
for each $j = 1, \dotsc, m_2$. since the GCM of $R_{F_n, G_n}$ is with probability tending to one piecewise linear with knots at the $y_j$ and $\theta_n^* = \partial_- \n{GCM}_{[0,1]}( R_{F_n, G_n}) \circ G_n$, we then have that with probability tending to one that $\theta_n^*$ is a left-continuous step function with jumps at the $y_j$. Also, since $G_n(z) = 0$ for $z < y_1$ and $R_{F_n, G_n}(u) = 0$ for all $u \leq 0$,  $\theta_n^*(z) = 0$ for $z < y_1$.

We now have
\begin{align*}
 n^{1/2}[\theta_n^*(y_j) - \theta_0(y_j) ]  &= n^{1/2}\left[  \frac{F_n(y_j) - F_n(y_{j-1})}{G_n(y_j) - G_n(y_{j-1})} - \frac{F_0(y_j) - F_0(y_{j-1})}{G_0(y_j) - G_0(y_{j-1})} \right].
 \end{align*}
 Using the notation introduced in Section~\ref{sec:iso}, this can be written as
 \begin{align*}
 &n^{1/2}\left\{  \frac{\left[ \frac{1}{n} \sum_{i=1}^n W_{i1} \right] /\left[ \frac{1}{n} \sum_{i=1}^n W_{i2} \right]}{ \left[ \frac{1}{n} \sum_{i=1}^nW_{i3} \right] /\left[ \frac{1}{n} \sum_{i=1}^n W_{i4} \right]}-   \frac{ E_0(W_{i1}) / E_0 (W_{i2})}{  E_0(W_{i3}) / E_0 (W_{i4}) }  \right\} &= n^{1/2}\left\{  g \left( \frac{1}{n} \sum_{i=1}^n W_i\right) -   g\left( E_0(W_i) \right)  \right\},
 \end{align*}
 where $W_i = (W_i1, \dotsc, W_{i4})^T$ for $W_{i1} = I(D_i = 1, y_{j-1} < Z_i \leq y_j)$, $W_{i2} = I(D_i = 1)$, $W_{i3}= I(D_i = 0, y_{j-1} < Z_i \leq y_j)$, and $W_{i4} = I(D_i = 0)$, and $g(w_1, w_2, w_3, w_4) = \frac{w_1 / w_2}{w_3 / w_4}$. By the Central Limit Theorem, 
 \[\sqrt{n} \left\{  \frac{1}{n} \sum_{i=1}^n W_{i} -  E_0(W_i) \right\} \indist N_4(0, V_0),\]
  where the $(j,k)$ element of the covariance matrix $V_0$ equals $E_0(W_j W_k) - E_0(W_j) E_0(W_k)$.  Applying the delta method to the function $g$ yields (after some algebra) $n^{1/2}[\theta_n^*(y_j) - \theta_0(y_j) ]  \indist N\left(0, \sigma_0^2(y_j)\right)$, where $ \sigma_0^2(y_j)$ equals
  \begin{align*}
\theta_0(y_j) \frac{\pi_0 [ F_0(y_j) - F_0(y_{j-1})] + (1-\pi_0) \Delta G_0(y_j) - [ F_0(y_j) - F_0(y_{j-1})]\Delta G_0(y_j)}{\pi_0(1-\pi_0) [\Delta G_0(y_j)]^2}.
 \end{align*}
\end{proof}

\begin{proof}[\bfseries Proof of Theorem~\ref{thm:cont_dist_fns}]
We note that  $f_0(z) / g_0(z) \leq f_0(z') / g_0(z')$ for all $z < z'$ in $[a,b]$ implies that 
\[\pi_0^2 f_0(z) f_0(z') + \pi_0(1-\pi_0) f_0(z) g_0(z') \leq \pi_0^2 f_0(z) f_0(z') + \pi_0(1-\pi_0) f_0(z') g_0(z),\]
which implies that $z \mapsto \pi_0 f_0(z) / [\pi_0f_0(z) + (1-\pi_0) g_0(z)]$ is non-decreasing on $[a,b]$. Therefore,
\begin{align*}
( G_0 \circ H_0^{-1})' &= \frac{g_0 \circ H_0^{-1}}{\pi_0f_0 \circ H_0^{-1}+ (1-\pi_0) g_0\circ H_0^{-1}}\\
& = \frac{1}{1-\pi_0} \left[ 1 -  \frac{\pi_0 f_0\circ H_0^{-1} }{\pi_0f_0\circ H_0^{-1}  + (1-\pi_0) g_0\circ H_0^{-1} } \right]
\end{align*}
is non-increasing on $H_0([a,b]) = [0,1]$. Hence, $G_0 \circ H_0^{-1}$ is concave on $[0,1]$, so 
\[\n{LCM}_{[0,1]}(G_0 \circ H_0^{-1}) \circ H_0 = G_0 \circ H_0^{-1} \circ H_0 = G_0.\]

We now note that since $G_n \circ H_n^{-}(u) \geq G_n(y_{m_2}) = 1$ for any $u \geq h_{m_2}$, $\n{LCM}_{[0,h_{m_2}]}( G_n \circ H_n^-) = \n{LCM}_{[0,1]}( G_n \circ H_n^-)$. Furthermore, since $G_n^*$ only jumps at $y_1, \dotsc, y_{m_2}$, we have
\[ G_n^*(y) = \n{LCM}_{[0,1]}( G_n \circ \tilde{H}_n^-) \circ \tilde{H}_n(y)\]
for any $y \in \d{R}$, where $\tilde{H}_n := \pi_n F_n \circ L_n + (1-\pi_n) G_n$ for $L_n(z) := G_n^- \circ G_n(z) = \max\{Y_j : Y_j \leq z\}$. 

Using the notation of Section~\ref{sec:iso}, we can write
\[ \pi_n F_n(x) = \frac{n_1}{n} \frac{1}{n_1} \sum_{i=1}^n D_i I(Z_i \leq x) = \d{P}_n \omega_x\]
for $\omega_x(d,x) := d I(z \leq x)$, and similarly $(1-\pi_n) G_n(y) =  \d{P}_n \eta_y$ for $\eta_y(d,z) := (1-d) I(z \leq y)$. We also have $P_0 \omega_x = \pi_0 F_0(x)$ and $P_0 \eta_y = (1-\pi_0) G_0(y)$. By standard empirical process theory, we therefore have that 
\[\{n^{1/2}[\pi_n F_n(x) - \pi_0 F_0(x)] : x \in \d{R}\} = \{n^{1/2} (\d{P}_n - P_0) \omega_x : x \in \d{R}\}\]
 and 
 \[\{n^{1/2}[(1-\pi_n) G_n(y) - (1-\pi_0) G_0(y)] : y \in \d{R}\} = \{n^{1/2} (\d{P}_n - P_0) \eta_y : y \in \d{R}\}\] 
 converge  weakly (jointly) as processes indexed by $\ell^\infty(\d{R})$ to 
 \[(\d{G}_1 \circ [\pi_0F_0], \d{G}_2 \circ [(1-\pi_0)G_0])\]
  for $\d{G}_1$ and $\d{G}_2$ independent Brownian bridge processes. The two processes are independent because the covariance between the processes is easily seen to be zero. Since the density of $G_0$ is bounded strictly away from zero on $[a, b]$, $n^{1/2}( [(1-\pi_n)G_n]^- - [(1-\pi_0)G_0]^{-1})$ converges weakly in $\ell^\infty(0,1)$ to $-\d{G}_2 / [(1-\pi_0)g_0] \circ [\pi_0G_0]^{-1}$  by Lemma 3.9.23 of \citealp{van1996weak}. Hence, by Hadamard differentiability of the composition map (see Lemma  3.9.27 of \citealp{van1996weak}), the functional delta method yields
  \[ n^{1/2}( G_n^- \circ G_n - \n{Id}) = n^{1/2}( [(1-\pi_n)G_n]^- \circ [\pi_nG_n] - \n{Id})\]
  converges weakly in $\ell^\infty[a, b]$ to
\begin{align*}
&-(\d{G}_2 \circ [(1-\pi_0)G_0]) / [(1-\pi_0)g_0 \circ [(1-\pi_0)G_0]^{-1} \circ [(1-\pi_0)G_0])\\
&\qquad + (\d{G}_2 \circ [(1-\pi_0)G_0]) / ([(1-\pi_0)g_0 \circ [(1-\pi_0)G_0]^{-1} \circ [(1-\pi_0)G_0]) = 0,
\end{align*}
so that $\sup_{z \in [ a, b]} | L_n(z) - z| = \fasterthan(n^{-1/2})$. Hence, $n^{1/2}(\pi_n F_n \circ L_n - \pi_0 F_0)$ converges weakly to $\d{G}_1 \circ [\pi_0F_0]$ in $\ell^\infty[ a, b]$, and so $n^{1/2}(\tilde{H}_n - H_0)$ converges weakly to $\d{G}_1 \circ [\pi_0 F_0] + \d{G}_2 \circ [(1-\pi_0) G_0]$ in $\ell^\infty[ a, b]$. Since $G_0$ and $F_0$ are both continuously differentiable on $[a,b]$, so is $H_0$, and since the derivative of $G_0$ is bounded away from zero, so is the derivative of $H_0$. Therefore, using Lemma 3.9.23 of \citealp{van1996weak} again, $n^{1/2}(\tilde{H}_n^- -H_0^-)$ converges weakly in $\ell^\infty(0,1-\varepsilon)$ to $(\d{G}_1 \circ [\pi_0 F_0] + \d{G}_2 \circ [(1-\pi_0) G_0]) \circ H_0^{-1} / (h_0 \circ H_0^{-1})$ for any $\varepsilon > 0$, where $h_0 := H_0' = \pi_0 f_0 + (1-\pi_0) g_0$.  Then, using the functional delta method for composition again, we have that $n^{1/2}[G_n \circ \tilde{H}_n^- - G_0 \circ H_0^{-1}]$ converges weakly to
\[ \left[1 -   \frac{g_0 \circ H_0^{-1}}{h_0 \circ H_0^{-1}}  \right] \d{G}_2 \circ [(1-\pi_0) G_0] \circ H_0^{-1}  - \frac{g_0 \circ H_0^{-1}}{h_0 \circ H_0^{-1}}  \d{G}_1 \circ (\pi_0 F_0) \circ H_0^{-1},\]
which we define as $\d{G}_3$. Now by Proposition~2.1 of \cite{beare2017weak}, 
\[ n^{1/2}[\n{LCM}_{[0,1]}(G_n \circ \tilde{H}_n^-) - \n{LCM}_{[0,1]}(G_0 \circ H_0^{-1})]\]
converges weakly to $\n{LCM}_{[0,1], G_0 \circ H_0^{-1}}'(\d{G}_3)$. Using Hadamard differentiability of composition once more, we have that
\[ n^{1/2}[G_n^* - G_0] =  n^{1/2}[\n{LCM}_{[0,1]}(G_n \circ \tilde{H}_n^-) \circ \tilde{H}_n - \n{LCM}_{[0,1]}(G_0 \circ H_0^{-1}) \circ H_0]\]
converges weakly to
\[ \n{LCM}_{[0,1], G_0 \circ H_0^{-1}}'(\d{G}_3) \circ H_0 + \frac{g_0}{h_0}\left[ \d{G}_1 \circ (\pi_0 F_0) + \d{G}_2 \circ [(1-\pi_0) G_0] \right]\]
If $f_0 / g_0$ is strictly increasing on $[a, b]$, then $G_0 \circ H_0^{-1}$ is strictly concave on $[0,1]$, in which case $ \n{LCM}_{[0,1], G_0 \circ H_0^{-1}}'$ is the identity operator by Proposition~2.2 of \cite{beare2017weak}. Hence, in this case $ n^{1/2}[G_n^* - G_0]$ converges weakly to 
\begin{align*}
& \d{G}_3 \circ H_0 +  \frac{g_0}{h_0}\left[ \d{G}_1 \circ (\pi_0 F_0) + \d{G}_2 \circ [(1-\pi_0) G_0] \right] \\
&\qquad = \left[1 -   \frac{g_0}{h_0}  \right] \d{G}_2 \circ [(1-\pi_0) G_0]  - \frac{g_0}{h_0}  \d{G}_1 \circ (\pi_0 F_0) \\
&\qquad\qquad  +  \frac{g_0}{h_0}\left[ \d{G}_1 \circ (\pi_0 F_0) + \d{G}_2 \circ [(1-\pi_0) G_0] \right] \\
&\qquad = \d{G}_2 \circ [(1-\pi_0) G_0], 
\end{align*}
which, as noted above, is the same limit distribution as $n^{1/2}[ G_n - G_0]$. Furthermore, we have
\begin{align*}
n^{1/2}\|G_n^* - G_n\|_{\infty} &\leq n^{1/2}\|[\n{LCM}_{[0,1]}(G_n \circ \tilde{H}_n^-) \circ \tilde{H}_n - G_n \circ \tilde{H}_n^- \circ \tilde{H}_n \|_{\infty} \\
&\qquad + n^{1/2}\|G_n \circ \tilde{H}_n^- \circ \tilde{H}_n - G_n\|_{\infty}.
\end{align*}
When $f_0 / g_0$ is strictly increasing so that $ \n{LCM}_{[0,1], G_0 \circ H_0^{-1}}'$ is the identity, the functional delta method (e.g.\ Theorem~3.9.4 of \citealp{van1996weak}) implies that
\begin{align*}
& n^{1/2}\|[\n{LCM}_{[0,1]}(G_n \circ \tilde{H}_n^-) \circ \tilde{H}_n - G_n \circ \tilde{H}_n^- \circ \tilde{H}_n \|_{\infty}\\
&\qquad \leq  n^{1/2}\|[\n{LCM}_{[0,1]}(G_n \circ \tilde{H}_n^-)  - G_n \circ \tilde{H}_n^-  \|_{\infty} = \fasterthan(1).
 \end{align*}
Similarly, since as shown above, $n^{1/2}(\tilde{H}_n^- \circ \tilde{H}_n - \n{Id})$ converges weakly in $[a,b]$ to 0, $n^{1/2}\|G_n \circ \tilde{H}_n^- \circ \tilde{H}_n - G_n\|_{\infty} = \fasterthan(1)$. Therefore, $n^{1/2}\|G_n^* - G_n\|_{\infty} = \fasterthan(1)$ if $f_0 / g_0$ is strictly increasing on $[a, b]$.

Now we turn attention to $F_n^*$. By Theorem~\ref{thm:mle}, for each $y \in \{Y_1, \dotsc, Y_n\}$, we know that $F_n^*(y_k) = \n{GCM}_{[0,h_{m_2}]}(F_n \circ H_n^-) \circ H_n(y_k)$. We can extend the GCM operation to entirety of $[0,1]$, so that $F_n^*(y_k) = \n{GCM}_{[0,1]}(F_n \circ H_n^-) \circ H_n(y_k)$, because the slope of the secant of $F_n \circ H_n^-$ from $h_{m_2}$ to $H_n(x_j)$ for any $x_j > y_{m_2}$ is $[F_n(x_j) - F_n(y_{m_2})] / [H_n(x_j) - h_{m_2}] = 1 / \pi_n$, while the slope of the secant from any other $z$ in the support of $H_n$ is $[F_n(y_{m_2}) - F_n(z)]  /  [H_n(y_{m_2}) - H_n(z)] \leq [F_n(y_{m_2}) - F_n(z)]  /  [\pi_n \{F_n(y_{m_2}) - F_n(z)\}] = 1/\pi_n$. Therefore, performing the GCM over $[0,1]$ rather than $[0,h_{m_2}]$ cannot change the value of the GCM for any $u \leq h_{m_2}$.

We now define $F_n^\ell(y) := \n{GCM}_{[0,1]}(F_n \circ H_n^-) \circ H_n \circ L_n$, where $L_n = G_n^- \circ G_n$ as above, so that $\bar{F}_n$ is the right-continuous step function with jumps at $y_1, \dotsc, y_{m_2}$ and agreeing with $F_n^*$ at these points. We similarly define $F_n^u := \n{GCM}_{[0,1]}(F_n \circ H_n) \circ H_n \circ R_n$, where $R_n := G_n^- \circ \bar{G}_n$ for $\bar{G}_n(y) := \frac{1}{n}\sum_{i=1}^n I(Y_i < y) + 1/n$. Since the $Y_j$'s are unique with probability one, $\bar{G}_n$ is a left-continuous version of $G_n$ that agrees at $y_1, \dotsc, y_{m_2}$, and $\bar{G}_n \geq G_n$. Therefore, since any MLE $F_n^*$ is a proper CDF, we have $F_n^\ell \leq F_n^* \leq F_n^u$. One can show that $\|F_n^\ell - F_n\|_{\infty} = \fasterthan(n^{-1/2})$ and $\|F_n^u - F_n\|_{\infty} = \fasterthan(n^{-1/2})$ when $f_0 / g_0$ is strictly increasing using the same argument as that used above for showing that $\|G_n^* - G_n\| = \fasterthan(n^{-1/2})$. We then have $\|F_n^* - F_n\|_{\infty} = \fasterthan(n^{-1/2})$ as well.
\end{proof}

\begin{proof}[\bfseries Proof of Theorem~\ref{thm:cons}]
The conditions of Theorem~1 of \cite{westling2019unified} are satisfied by the uniform consistency of empirical distribution functions.
\end{proof}

\begin{proof}[\bfseries Proof of Theorem~\ref{thm:dist}]
This result follows by the delta method, as discussed in the text.
\end{proof}

\clearpage

\section*{Additional simulations: discrete case}

We now present results from a numerical study of the properties of the maximum likelihood estimator in the case where both $F_0$ and $G_0$ are fully discrete. We set $F_0$ and $G_0$ as the distribution functions of Poisson random variables with rates 6 and 4, respectively, and we set $\pi_0$ to $0.4$. We simulated 1000 datasets each for $n \in \{500, 1000, 5000, 10000\}$ and estimated the maximum likelihood estimator $\theta_n^*$, the empirical mass ratio function, defined as the ratio of the empirical mass functions of $X_1, \dotsc, X_{n_1}$ and $Y_1, \dotsc, Y_{n_2}$, and the sample splitting estimators with $m \in \{5, 10 ,20\}$ \citep{banerjee2019divide}. We computed Wald-type confidence intervals (constructed around $\log \theta_n^*$ and exponentiated) using the asymptotic variance provided in Section~4.1 of the main text, likelihood ratio-based confidence intervals, and confidence intervals around the sample splitting estimators as outlined in Section~5 of the main text. 

\begin{figure}[ht!]
\centering
\includegraphics[width=.49\linewidth]{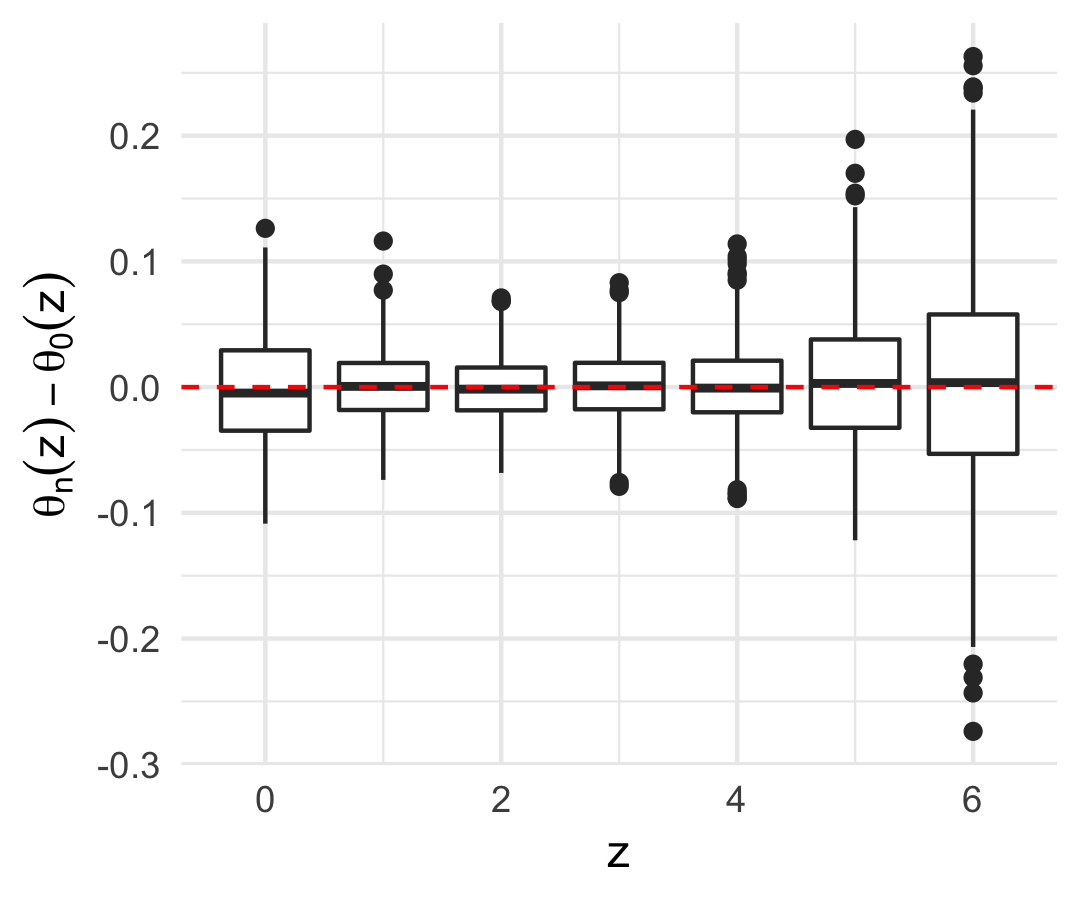}
\includegraphics[width=.49\linewidth]{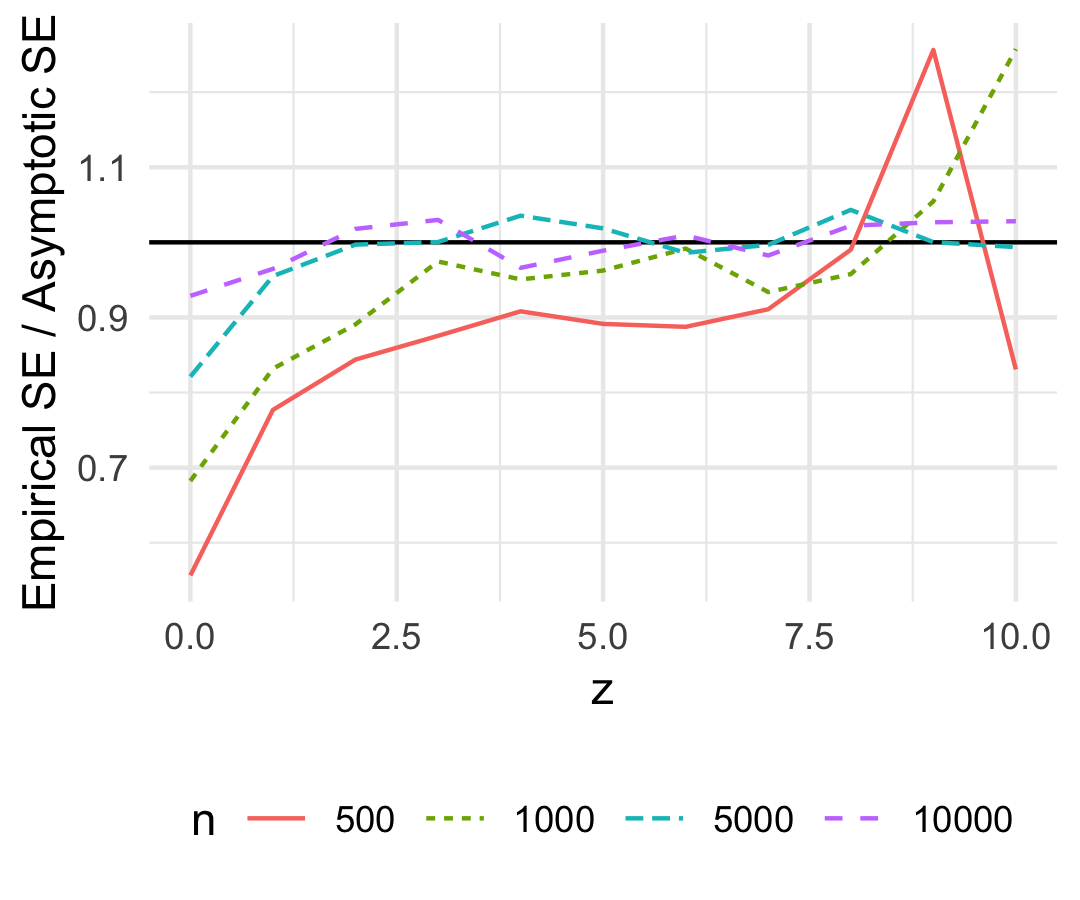}
\caption{Left: boxplots of $\theta_n^*(z) - \theta_0(z)$ with $n=10K$ in the fully discrete case. Right: empirical standard errors of $n^{1/2}[\theta_n^*(z) - \theta_0(z)]$ divided by the limit theory-based counterparts for $z \in \{0,1, \dotsc, 10\}$.}
\label{fig:sd_empirical_vs_true_disc}
\end{figure}

The left panel of Figure~\ref{fig:sd_empirical_vs_true_disc} displays the distribution of $\theta_n^*(z) - \theta_0(z)$ for $z \in \{0,1, \dotsc, 6\}$, and  demonstrates that $\theta_n^*$ is approximately unbiased in large samples. The right panel of Figure~\ref{fig:sd_empirical_vs_true_disc} displays the ratio of the empirical standard deviation of $n^{1/2}[ \theta_n^*(z) - \theta_0(z)]$ to the standard deviation based on the asymptotic theory, and demonstrates that the empirical standard deviation of $\theta_n^*(z)$ approaches the standard deviation defined by the limit theory as the sample size grows, and that $\theta_n^*(z)$ is more efficient than the limit theory suggests in smaller samples for small values of $z$.

\begin{figure}[ht!]
\centering
\includegraphics[width=\linewidth]{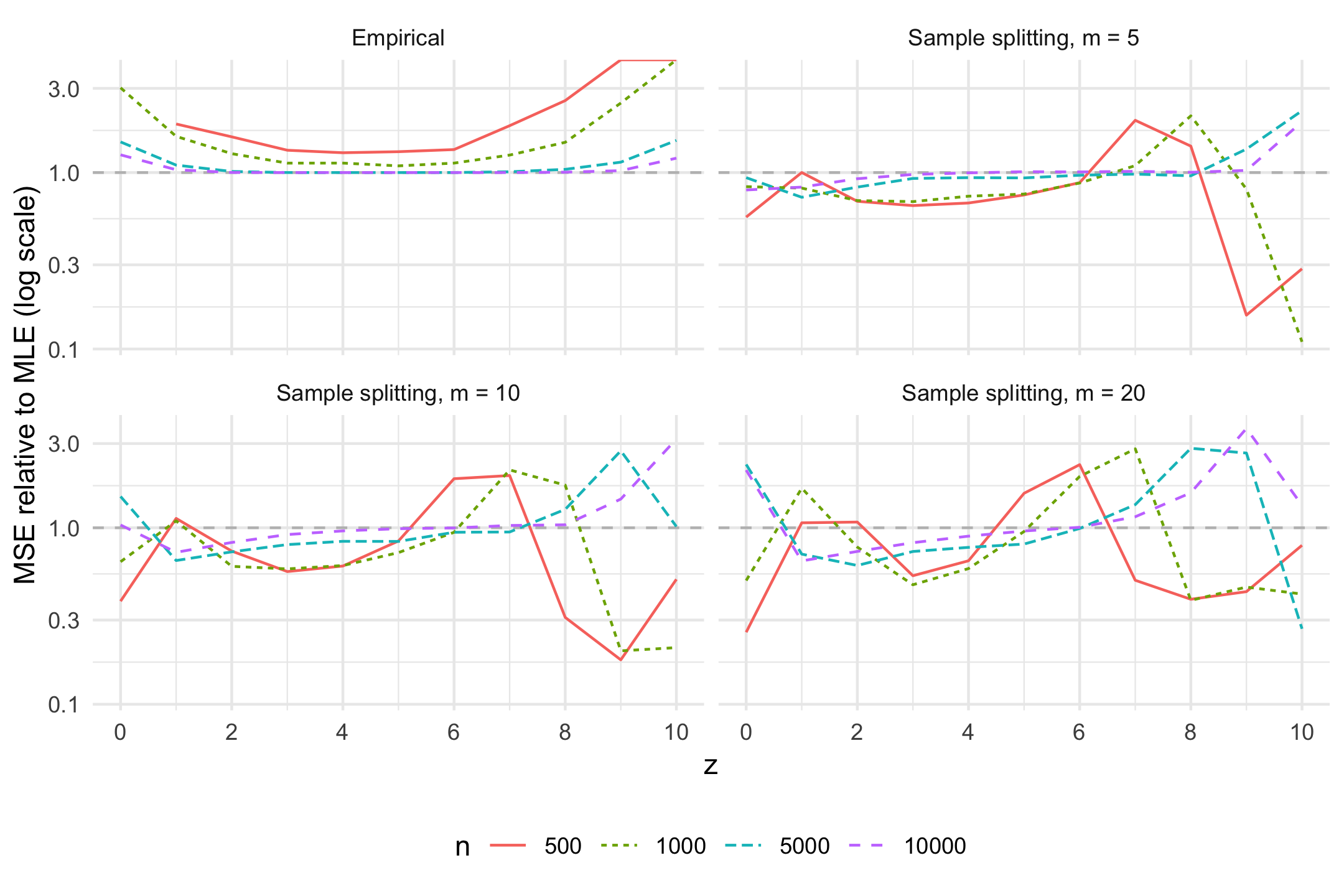}
\caption{Relative mean squared errors of the empirical estimator and the sample splitting estimators to the maximum likelihood estimator for $z \in \{0,1, \dotsc, 10\}$ and various sample sizes $n$ in the fully discrete case. The maximum likelihood has better mean squared error for $y$-values greater than one, and the other estimator has better mean squared error for $y$-values less than one.}
\label{fig:relative_mses_disc}
\end{figure}

Figure~\ref{fig:relative_mses_disc} displays the ratio of the mean squared errors of the empirical and sample splitting estimators to that of the maximum likelihood estimator. For the empirical estimator, this ratio approaches one as sample size grows, which agrees with our theoretical result suggesting that the two estimators are asymptotically equivalent. However, in small samples, the maximum likelihood estimator has strictly smaller mean squared error than the empirical estimator. The mean squared errors of the sample splitting estimators also approach that of the maximum likelihood estimator as the sample size grows, which is concurrent with existing theory for $n^{-1/2}$-rate asymptotics.

\begin{figure}[ht!]
\centering
\includegraphics[width=\linewidth]{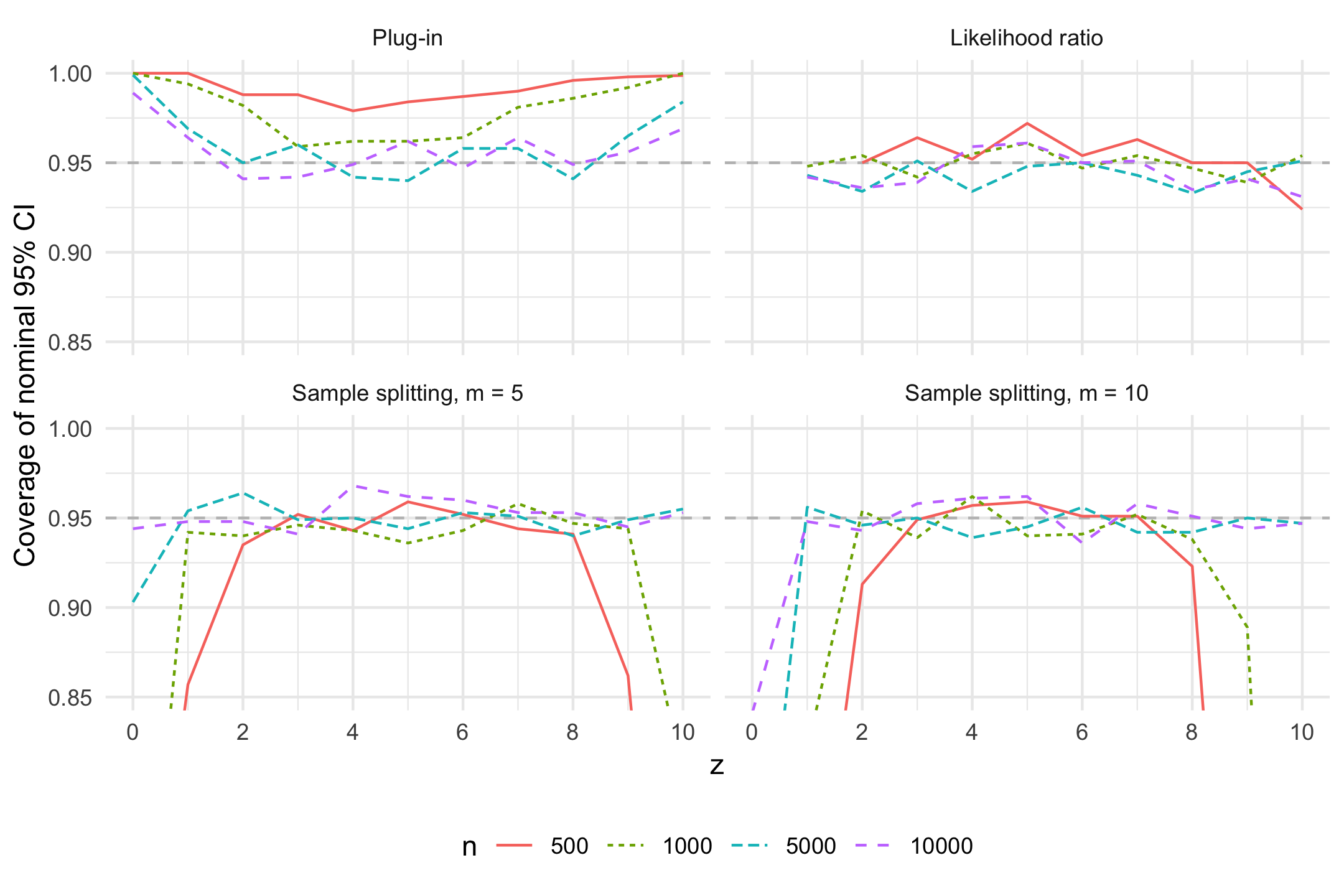}
\caption{Coverage of 95\% CIs in the fully discrete case for $z \in \{0,1,\dotsc, 10\}$, various sample sizes $n$, and four methods: the plug-in method centered around the log of the maximum likelihood estimator (upper left), the inverted likelihood ratio tests (upper right), and the sample splitting method with $m=5$ (lower left) and $m=10$ (lower right). Note that the likelihood ratio method does not provide intervals at the endpoints.}
\label{fig:coverages_disc}
\end{figure}

Figure~\ref{fig:coverages_disc} shows the empirical coverage of 95\% confidence intervals for $\theta_0(z)$ constructed using Wald-type confidence intervals with a plug-in standard error according to the results presented in Section~4.1 of the main text, the inverted likelihood ratio test approach of \cite{banerjee2001ratio}, and the sample splitting approach of \cite{banerjee2019divide} described in the main text. We note that the likelihood ratio approach does not provide intervals at the end point $z = 0$. The plug-in method is conservative in small samples, but its coverage approaches 95\% for $z \neq 0$ as $n$ grows. The likelihood ratio method provides excellent coverage at all sample sizes.  The sample splitting method has good coverage in large enough sample sizes. 

\clearpage

\subsection*{Additional simulations: continuous case}

We now present results from a numerical study of the properties of the maximum likelihood estimator in the case where both $F_0$ and $G_0$ are fully continuous. We set $F_0$ and $G_0$ as the distribution functions of exponential random variables with rates 1 and 2, respectively, and we set $\pi_0$ to $0.4$. We simulated 1000 datasets each for $n \in \{500, 1000, 5000, 10000\}$ and estimated the maximum likelihood estimator, the maximum smoothed likelihood estimator of \cite{yu2017ratio}, the non-monotone estimator based on kernel density estimates for each $z \in \{0, 0.1, \dotsc, 1.9, 2\}$, and the sample splitting estimator with $m \in \{5, 10 ,20\}$ \citep{banerjee2019divide}. We constructed confidence intervals at each $z$ using the transformed plug-in and likelihood ratio-based methods described in Section~4.2 of the main text.

\begin{figure}[ht!]
\centering
\includegraphics[width=.49\linewidth]{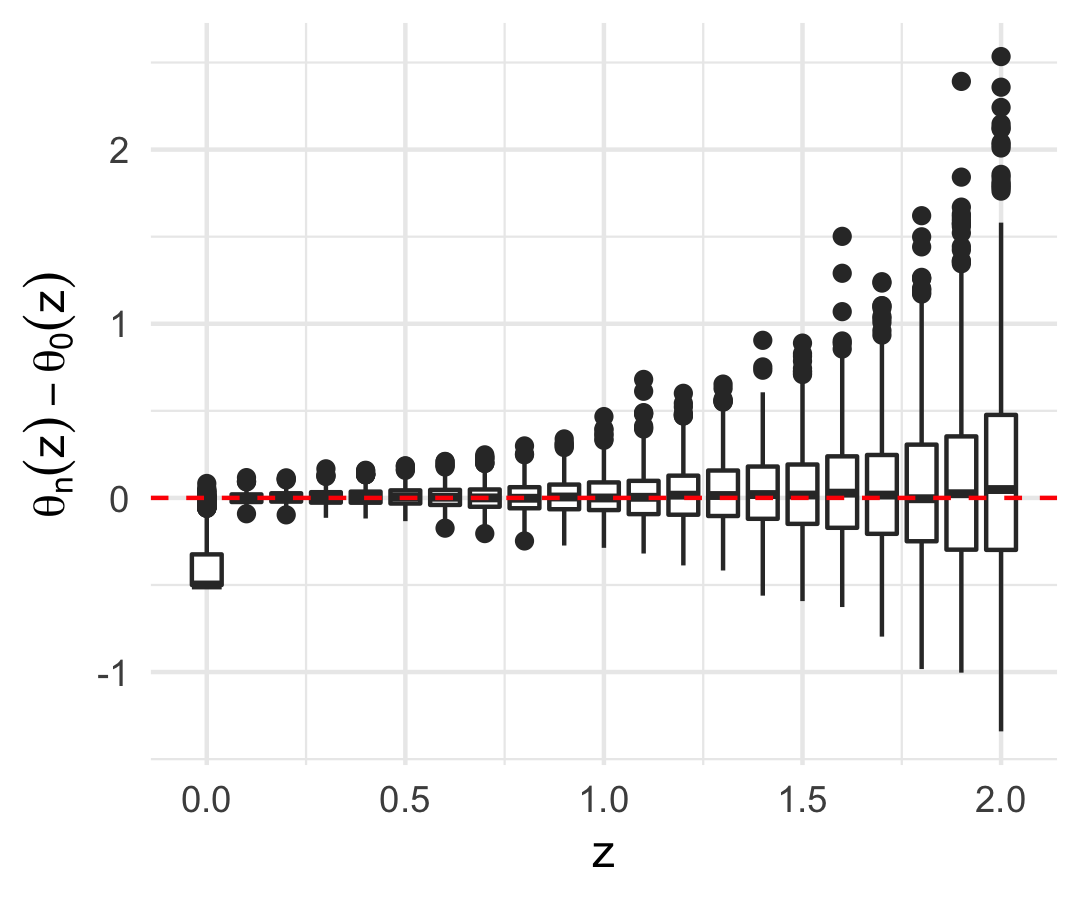}
\includegraphics[width=.49\linewidth]{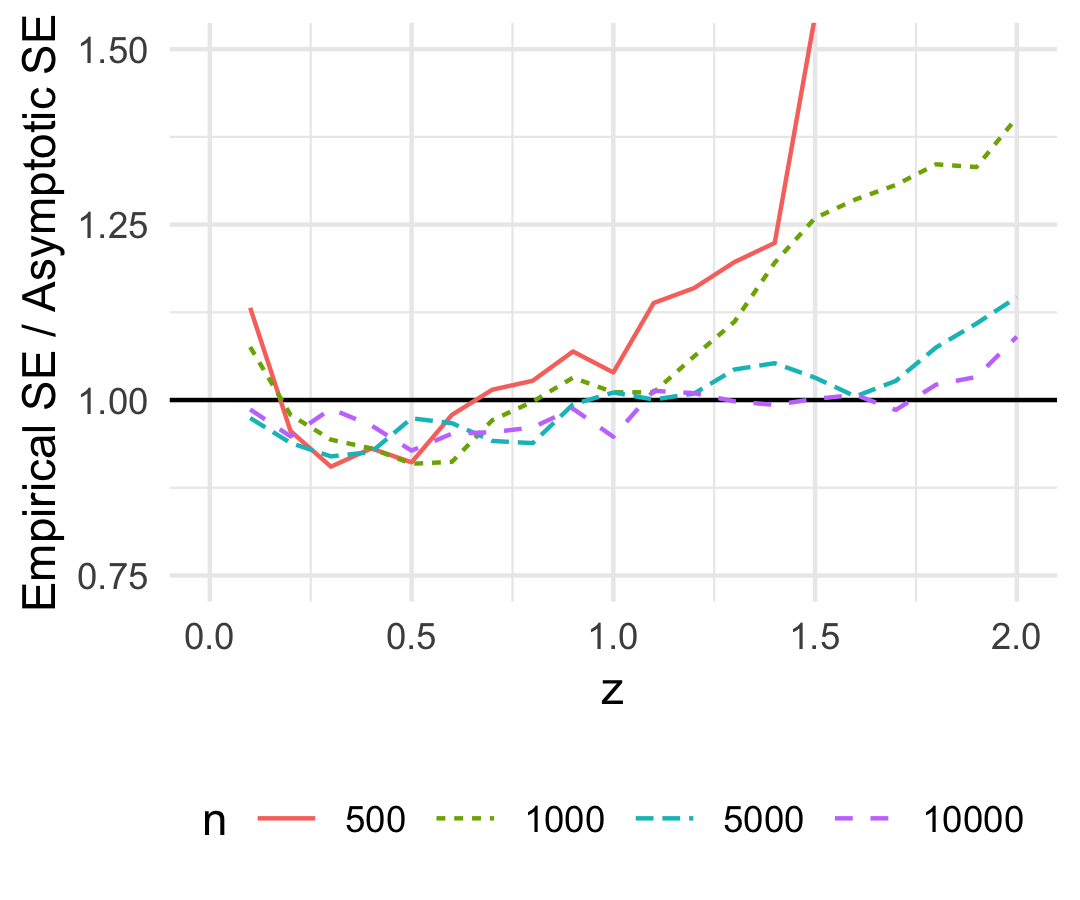}
\caption{Left: boxplots of $\theta_n^*(z) - \theta_0(z)$ with $n=10K$ in the fully continuous case. Right: empirical standard errors of $n^{1/2}[\theta_n^*(z) - \theta_0(z)]$ divided by the limit theory-based counterparts for $z \in [0,2]$.}
\label{fig:sd_empirical_vs_true_cont}
\end{figure}

The left panel of Figure~\ref{fig:sd_empirical_vs_true_cont} displays the distribution of $\theta_n^*(z) - \theta_0(z)$ for $z \in [0,2]$, and  demonstrates that the sampling distribution of $\theta_n^*$ is approximately centered around $\theta_0(z)$ in  large samples for $z > 0$. The right panel of Figure~\ref{fig:sd_empirical_vs_true_cont} displays the ratio of the empirical standard deviation of $n^{1/2}[ \theta_n^*(z) - \theta_0(z)]$ to the standard deviation based on the asymptotic theory, and demonstrates that the empirical standard deviation of $\theta_n^*(z)$ approaches the standard deviation defined by the limit theory as the sample size grows.

\begin{figure}[ht!]
\centering
\includegraphics[width=\linewidth]{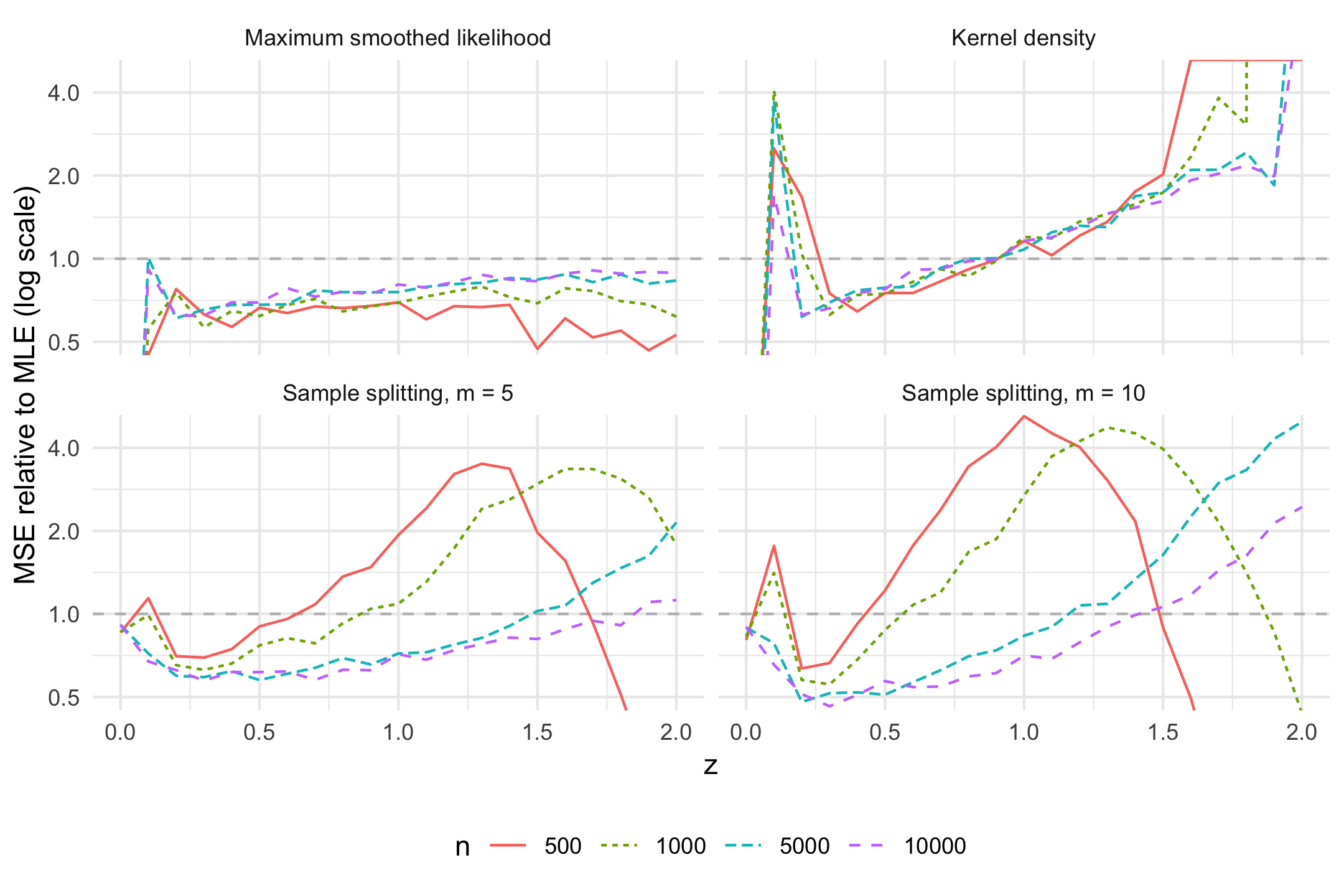}
\caption{Relative mean squared errors of the maximum smoothed likelihood estimator, the kernel density estimator, and the sample splitting estimators to the maximum likelihood estimator for $z \in [0,2]$ and various sample sizes $n$ in the fully continuous case. The maximum likelihood has better mean squared error for $y$-values greater than one, and the other estimator has better mean squared error for $y$-values less than one.}
\label{fig:relative_mses_cont}
\end{figure}

Figure~\ref{fig:relative_mses_cont} displays the ratio of the mean squared errors of maximum smoothed likelihood estimator, the kernel density estimator, and the sample splitting estimators to the maximum likelihood estimator. The maximum smoothed likelihood estimator is more efficient than the maximum likelihood estimator. The kernel density estimator is more efficient for some values of $z$, but less efficient for others. In large enough samples, the sample splitting estimators are more efficient than the maximum likelihood estimator, but in smaller samples, they are less efficient for some values of $z$. The sample size required for improvement grows with $m$, as does the gain in asymptotic efficiency.

\begin{figure}[ht!]
\centering
\includegraphics[width=\linewidth]{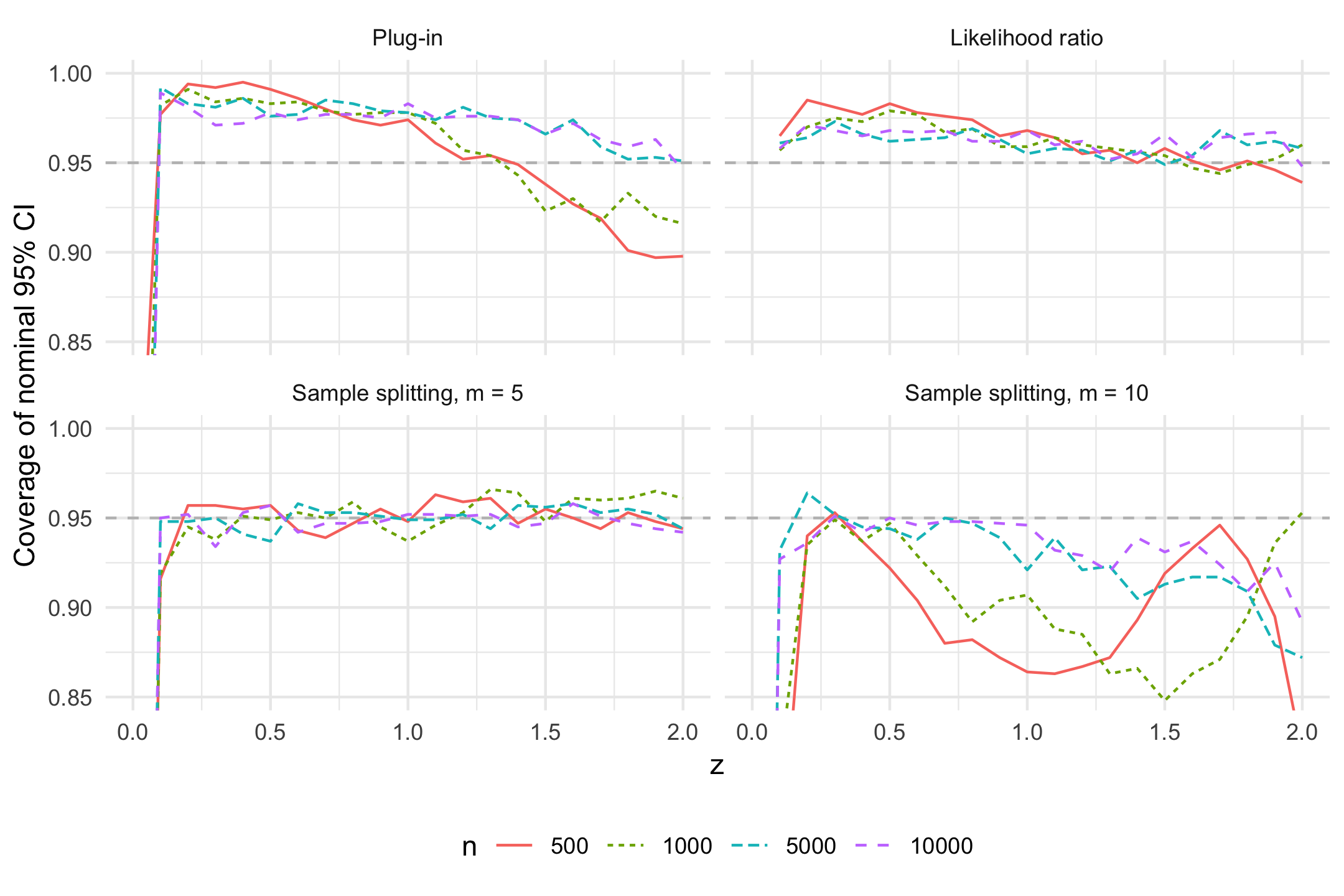}
\caption{Coverage of 95\% CIs in the fully continuous case for $z \in (0,2]$, various sample sizes $n$, and four methods: the plug-in method (upper left), the inverted likelihood ratio tests (upper right), and the sample splitting method with $m=5$ (lower left) and $m=10$ (lower right).}
\label{fig:coverages_cont}
\end{figure}

Finally, Figure~\ref{fig:coverages_cont} shows the empirical coverage of 95\% confidence intervals for $\theta_0(z)$ constructed using Wald-type confidence intervals with a plug-in standard error according to the results presented in Section~4.2 of the main text, the inverted likelihood ratio test approach of \cite{banerjee2001ratio}, and the sample splitting approach of \cite{banerjee2019divide} described in the main text. The plug-in method is conservative in large enough samples due to the difficulty of accurately estimating the derivative of $\theta_0$. The likelihood ratio method provides slightly conservative coverage at all sample sizes.  The sample splitting method has excellent coverage for $m=5$, but requires larger samples to have good coverage for $m=10$.

\clearpage

\subsection*{Additional simulations: flat case with jumps}

Here we present results from a numerical study of the properties of the various estimators in the case where $F_0$ and $G_0$ are mixed distributions, and $\theta_0 = dF_0/dG_0$ is discontinuous. We set $F_0 := (2/3) F_{0}^c + (1/3) \delta_0$, where $F_0^c$ is the uniform distribution on $[0,1]$ and $\delta_0$ is a discrete distribution with mass $1/6$ at $0$, $1/3$ at $1/2$, and $1/2$ at $1$. We set $G_0 := (2/3) F_0^c + (1/3)\gamma_0$, where $\gamma_0$ is a discrete distribution with mass $1/3$ each at 0, $1/2$, and $1$. We set $\pi_0$ to $0.4$. 

With these definitions, we have $\theta_0(x) = 1/2$ for $x = 0$, $\theta_0(x) = 1$ for $x \in (0,1)$, and $\theta_0(x) = 3/2$ for $x = 1$. Hence, $\theta_0$ has jumps at the extremal mass points $x = 0$ and $x = 1$, and is flat between these mass points. Therefore, our large-sample theory does not cover this case for two reasons: because $\theta_0$ is flat in the interior, and because it is discontinuous at the boundaries.

We simulated 1000 datasets each for $n \in \{500, 1000, 5000, 10000\}$ and estimated the maximum likelihood estimator, the maximum smoothed likelihood estimator of \cite{yu2017ratio}, the non-monotone estimator based on kernel density estimates for each $z \in \{0, 0.1, \dotsc, 1.9, 2\}$, and the sample splitting estimator with $m \in \{5, 10\}$ \citep{banerjee2019divide}. We constructed confidence intervals at each $z$ using the transformed plug-in and likelihood ratio-based methods described in Section~4.2 of the main text. We were unable to use the plug-in method of constructing confidence intervals because it failed in this case due to the difficulty of estimating the derivative of a flat function.

\begin{figure}[ht!]
\centering
\includegraphics[width=.49\linewidth]{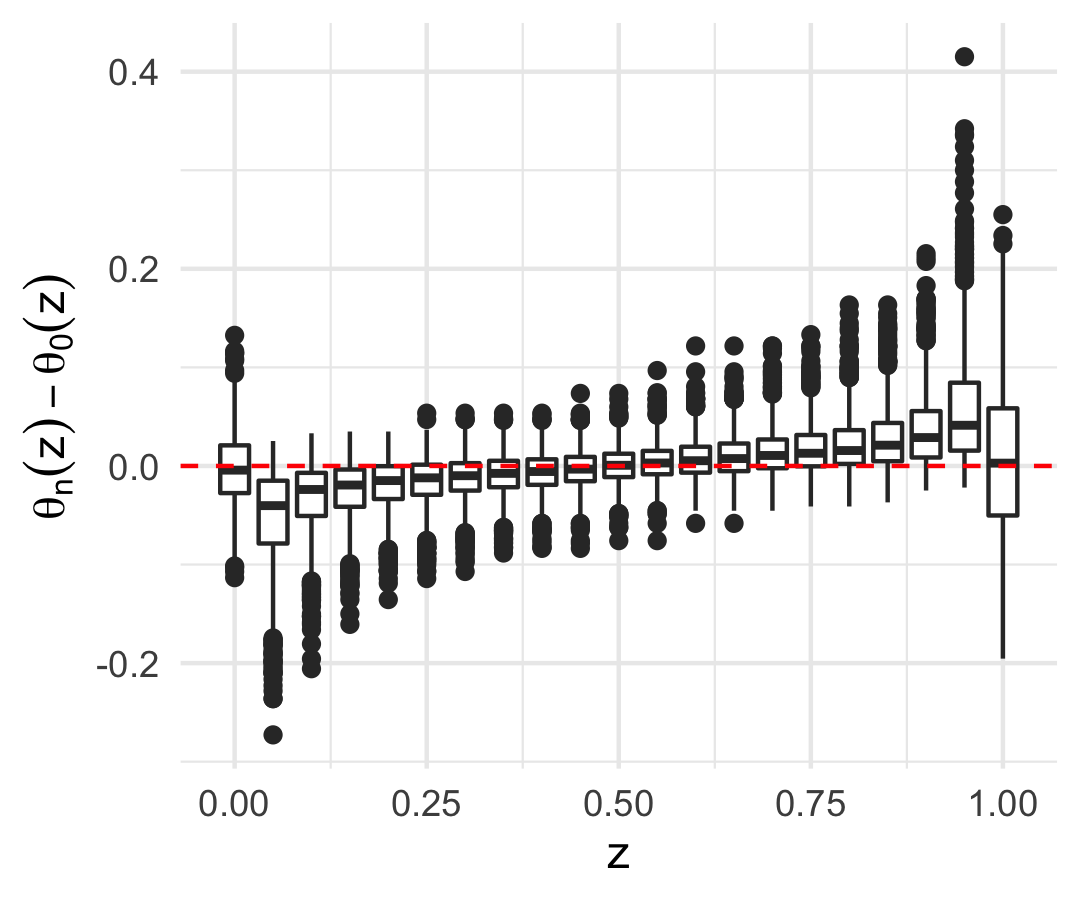}
\caption{Boxplots of $\theta_n^*(z) - \theta_0(z)$ with $n=10K$ in the flat case with jumps for $z \in [0,1]$.}
\label{fig:boxplots_flat_jumps}
\end{figure}

Figure~\ref{fig:boxplots_flat_jumps} displays the distribution of $\theta_n^*(z) - \theta_0(z)$ for $z \in [0,1]$. The pattern is quite interesting. For $z \in \{0,1\}$, the estimator appears to be centered around the truth. This suggests that the estimator may be consistent at mass points even if the function is discontinuous at these points or these points lie on the boundary of the domain. However, for $z \in (0, 0.25)$, the distribution of $\theta_n^*(z)$ is biased downward, and for $z \in (0.75, 1)$, the distribution is biased upward. This is likely due to the discontinuity of $\theta_0$ at 0 and 1: although the estimator is consistent for any $z \in (0,1)$, in any finite sample the estimator is flat in a region of the discontinuity, which biases the finite-sample distribution of the estimator near these discontinuities. We will see below that this also makes inference in these areas challenging.

\begin{figure}[ht!]
\centering
\includegraphics[width=\linewidth]{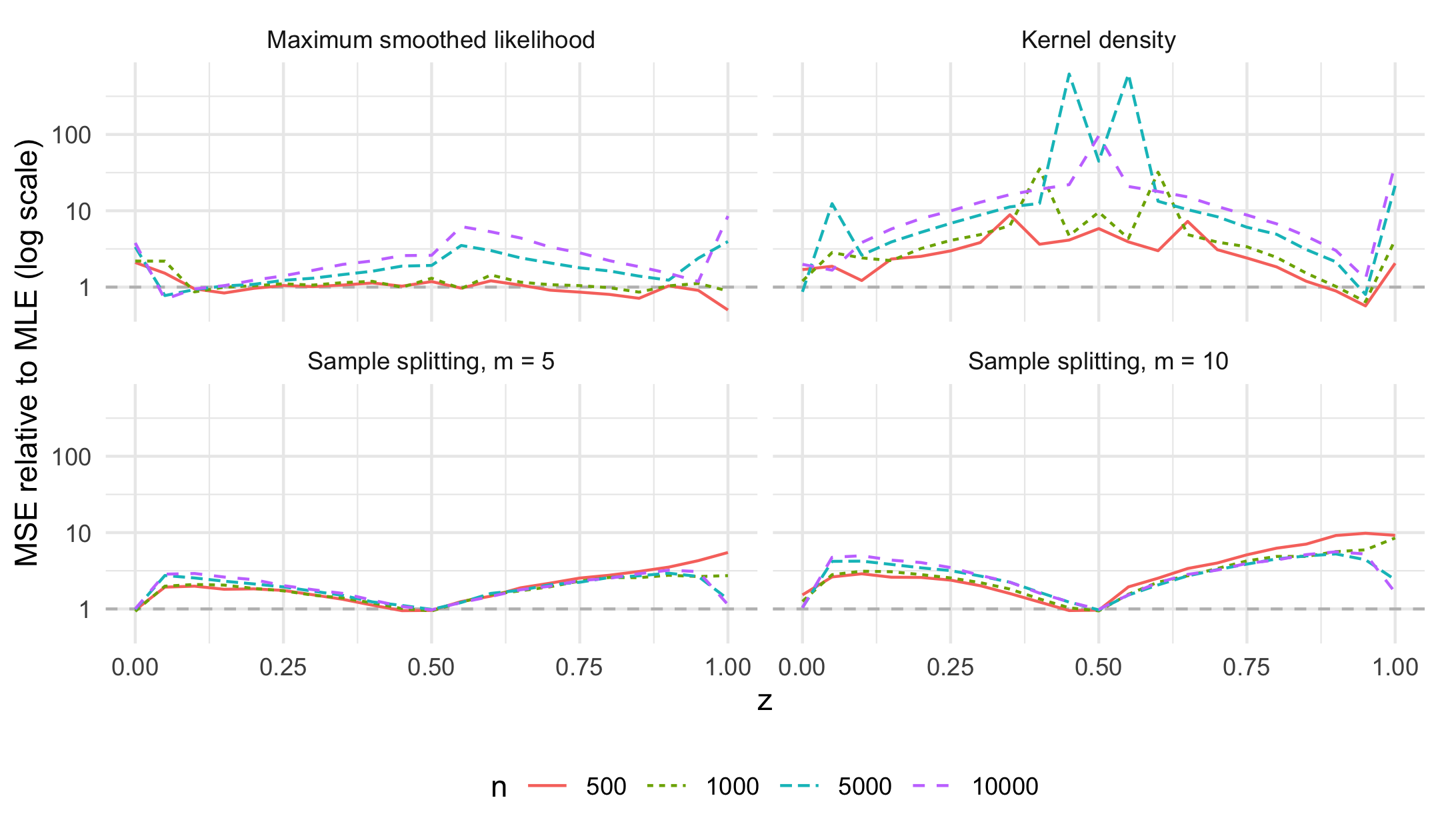}
\caption{Relative mean squared errors of the maximum smoothed likelihood estimator, the kernel density estimator, and the sample splitting estimators to the maximum likelihood estimator for $z \in [0,1]$ and various sample sizes $n$ in the flat case with jumps. .}
\label{fig:relative_mses_flat_jumps}
\end{figure}

Figure~\ref{fig:relative_mses_flat_jumps} displays the ratio of the mean squared errors of maximum smoothed likelihood estimator, the kernel density estimator, and the sample splitting estimators to the maximum likelihood estimator. The maximum smoothed likelihood estimator is comparable to the maximum likelihood estimator for $n \in \{500,1000\}$, but is less efficient for most $z$ in larger samples. This is especially true for $z \in \{0,1\}$, where the maximum likelihood estimator appears to benefit from the mass points.  The kernel density estimator is less efficient, and in large samples much less efficient, for all  $z$ except those very close to $0$ and $1$. Somewhat surprisingly, the sample splitting estimators are less efficient than the maximum likelihood estimator except for $z$ near the mass points. This is likely due to the fact that the sample splitting estimator inherit the bias of the maximum likelihood estimator at a smaller sample size, and the maximum likelihood estimator is biased near the points of discontinuity.

\begin{figure}[ht!]
\centering
\includegraphics[width=\linewidth]{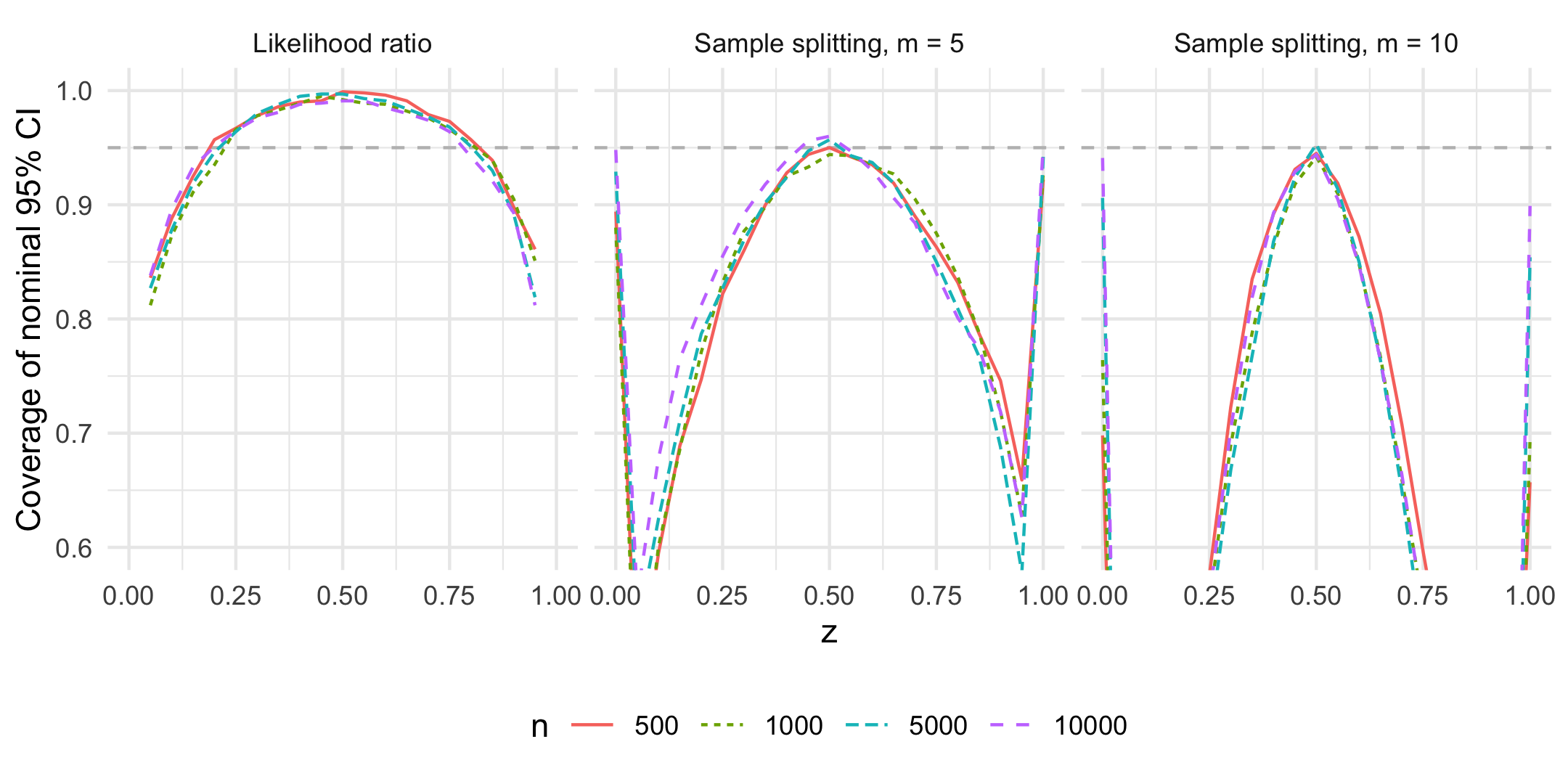}
\caption{Coverage of 95\% CIs in the flat case with jumps for $z \in [0,1]$, various sample sizes $n$, and three methods: the inverted likelihood ratio tests (left) and the sample splitting method with $m=5$ (middle) and $m=10$ (right).}
\label{fig:coverages_flat_jumps}
\end{figure}

Finally, Figure~\ref{fig:coverages_flat_jumps} shows the empirical coverage of 95\% confidence intervals for $\theta_0(z)$ constructed using the inverted likelihood ratio test approach of \cite{banerjee2001ratio} and the sample splitting approach of \cite{banerjee2019divide} described in the main text. None of the methods do well near $z \in \{0,1\}$ due to the bias of the estimators in these regions. The likelihood ratio method provides conservative coverage at all sample sizes for $z$ near $1/2$, which is because it relies on limit theory that only holds when $\theta_0$ is strictly increasing.  The sample splitting method has good coverage for $z$ at the mass points $\{0,1/2,1\}$, but poor coverage otherwise.

\clearpage

\section*{Additional data analysis results}

\begin{figure}[ht!]
\centering
\includegraphics[width=5in]{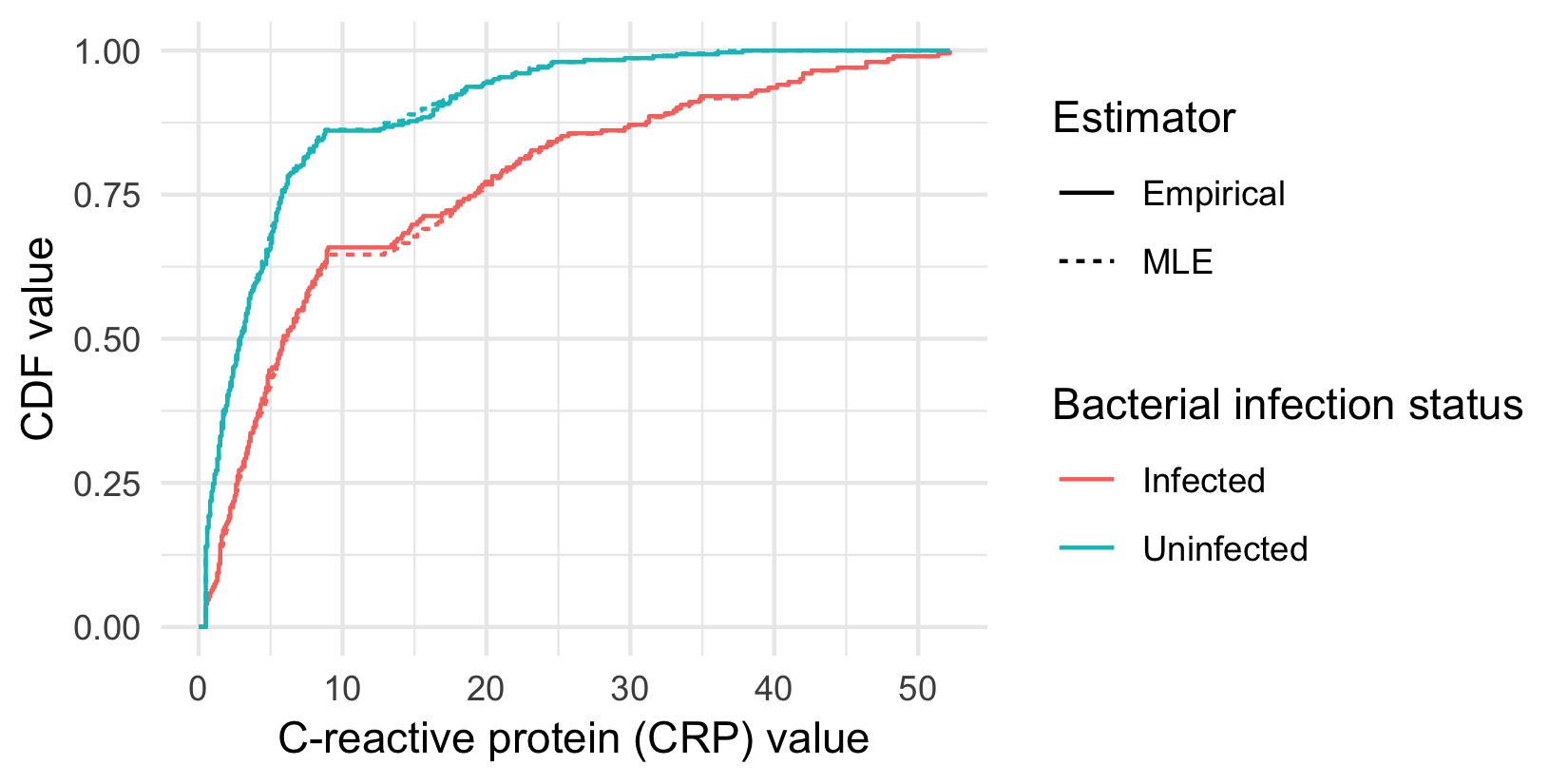}
\caption{Estimated cumulative distribution functions of C-reactive protein value among patients with bacterial infections and those without. Both the empirical distribution functions and the maximum likelihood estimators under the likelihood ratio order are shown.}
\label{fig:crp_cdfs}
\end{figure}

\begin{figure}[h]
\centering
\includegraphics[width=5in]{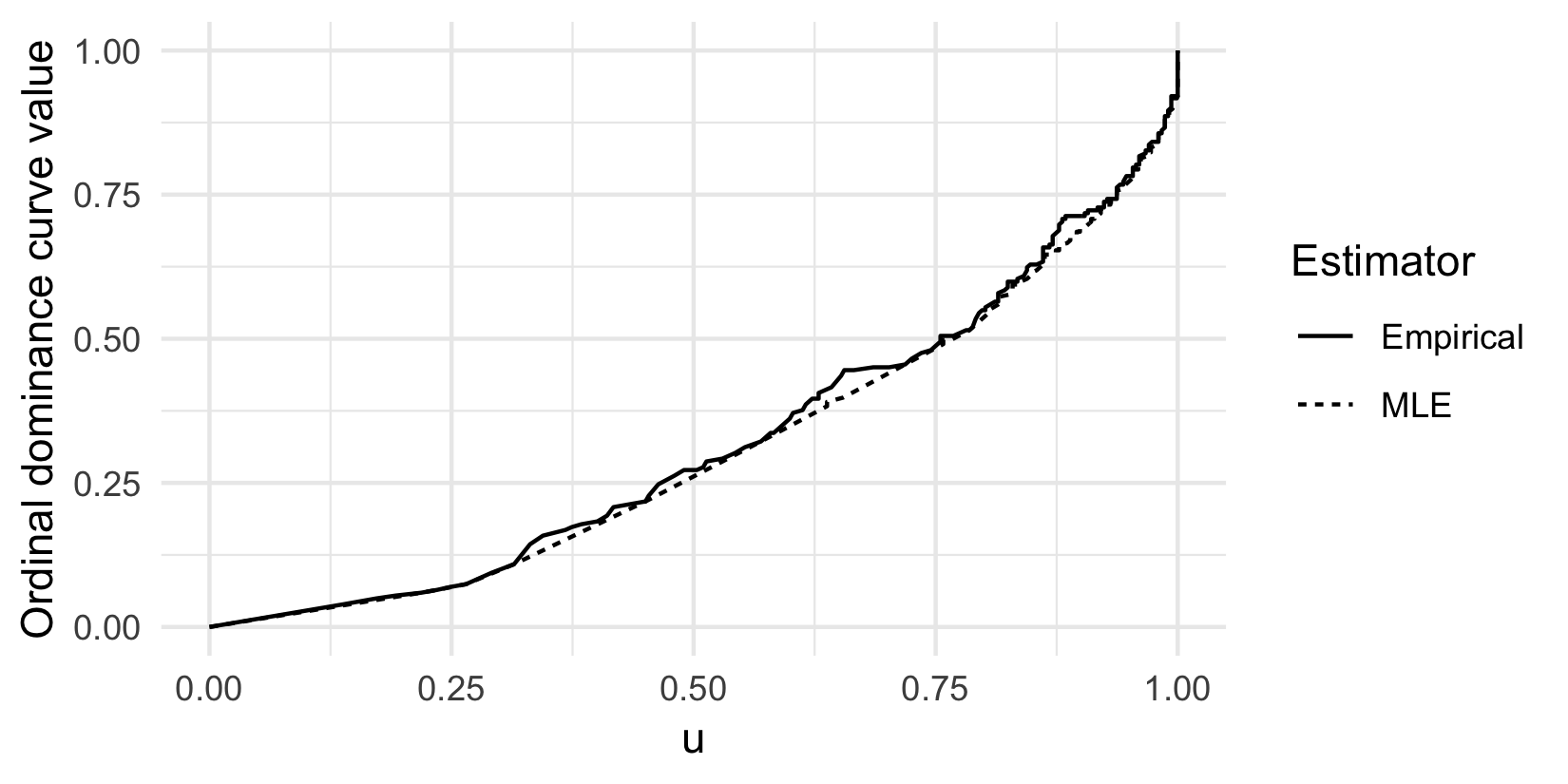}
\caption{Estimated ordinal dominance curve for C-reactive protein. Both the empirical distribution functions and the maximum likelihood estimators under the likelihood ratio order are shown.}
\label{fig:crp_odc}
\end{figure}

Figure~\ref{fig:crp_cdfs} displays the empirical and likelihood ratio order maximum likelihood cumulative distribution function estimates of C-reactive protein for patients with bacterial infections and those without.  Figure~\ref{fig:crp_odc} displays the empirical and likelihood ratio order maximum likelihood ordinal dominance curve estimates for C-reactive protein.

\end{document}